\documentclass[12pt,natbib,sort&compress]{iopart}
\usepackage{epsfig}
\usepackage[latin1]{inputenc}
\usepackage{graphicx}

\begin{document}

\title[Structure of $GeO_2$]{The structure of amorphous, crystalline and liquid $GeO_2$}

\author{M. Micoulaut$^{1}$, L. Cormier$^{2}$  and G.S. Henderson$^{3}$  
}
\address{}
\address{$^1$ Laboratoire de Physique Théorique de la Matière Condensée, 
Université Pierre  et Marie Curie,  CNRS UMR 7600,  Boite 121, 4 place
Jussieu, 75252 Paris Cedex 05, France}
\address{}
\address{$^2$ Institut de Minéralogie et de Physique des Milieux 
Condensés, Université Pierre et Marie Curie, Université Denis Diderot,
CNRS UMR 7590, 4 place Jussieu, 75252 Paris Cedex 05 France
}
\address{}
\address{$^3$ Department of Geology, University of Toronto, 22 Russell Street,
Toronto, Ontario M5S 3B1, Canada}
$$ $$
\begin{abstract}
Germanium   dioxide    ($GeO_2$)    is    a   chemical  analogue    of
$SiO_2$. Furthermore, it is also to some extent a structural analogue,
as the    low and high-pressure   short-range  order  (tetrahedral and
octahedral) is the same. However, a number  of differences exist.  For
example,  the $GeO_2$  phase  diagram  exhibits  a  smaller number  of
polymorphs, and all three  $GeO_2$ phases (crystalline, glass, liquid)
have an increased sensitivity to pressure, undergoing pressure induced
changes  at  much   lower   pressures than  their  equivalent  $SiO_2$
analogues.   In addition, differences  exist  in $GeO_2$ glass in  the
medium range order, resulting  in the glass transition temperature  of
germania being much lower than for silica.  This review highlights the
structure of amorphous $GeO_2$  by different experimental (e.g., Raman
and  NMR spectroscopy, neutron and  x-ray diffraction) and theoretical
methods    (e.g.,    classical  molecular      dynamics,  ab    initio
calculations). It also addresses the structure  of liquid
and    crystalline $GeO_2$ that   have  received much less attention. 
Furthermore, we  compare   and  contrast the structural    differences
between $GeO_2$ and $SiO_2$, as well as, along the $GeO_2-SiO_2$ join.
It is probably a very timely review as interest in this compound, that
can be investigated in the liquid state at relatively low temperatures
and pressures, continues to increase.
\end{abstract}



\maketitle
\tableofcontents

\section{Introduction}

Zachariasen   \cite{Zachariasen1928} proposed  the continuous   Random
network model (CRN) to explain the structure of  oxide glasses, and it
has subsequently  received wide acceptance  in describing glasses that
form continuous  random networks. To date the   majority of studies of
oxide  glasses have involved the investigation  of silica ($SiO_2$) or
borate ($B_2O_3$) glasses with, to a lesser extent, germania ($GeO_2$)
glasses. The structure of the latter has  generally been considered to
be comparable  to that of   silica glass despite differences  in  bond
lengths,  angles and      the   relative   size  of   $Ge$      versus
$Si$.    Experimental studies  of    amorphous  $GeO_2$ have generally
involved   either x-ray   or   neutron   scattering  and spectroscopic
techniques  such as x-ray   absorption spectroscopy (EXAFS/XANES)  and
Raman spectroscopy.  On the    other hand, theoretical   studies  have
generally employed classical or   {\em ab initio}  molecular  dynamics
calculations to gain insight into the structure of these materials. In
both approaches, the results of the studies  are often compared to the
known crystalline polymorphs of $GeO_2$. Here  we review the structure
of amorphous $GeO_2$ (glass and  liquid) from both an experimental and
theoretical  perspective, as well as,   comparing their structure with
that of amorphous $SiO_2$ (glass,  liquid). Furthermore, we review the
structure of    the  crystalline $GeO_2$  polymorphs,  both    at room
temperature and pressure and at elevated temperatures and pressures.

\section{Crystalline $GeO_2$ Polymorphs}

\subsection{Structure}
Crystalline $GeO_2$ exists   at ambient temperatures and pressures  as
one of two polymorphs (Figure  1): an $\alpha$-quartz-like  ($P3_221$)
trigonal        (hexagonal)    structure       \cite{Smith1964} or as a rutile-like tetragonal ($P4_2/mnm$) structure
\cite{Baur1971}.

\par

The $\alpha$-quartz-like  $GeO_2$ structure has been  shown  to be the
stable high temperature  phase \cite{Laubengayer1932}  and, while  the
structure is  very similar to  that of $\alpha$-quartz, there are some
distinct  differences. In particular the   $GeO_4$ tetrahedra are more
distorted due to greater  variation in the  $O-Ge-O$ angles within the
tetrahedron, which range from $106.3^o$ to  $113.1^o$ with a $Ge-O-Ge$
angle of $130.1^o$. This  is in contrast  to $\alpha$-quartz where the
$O-Si-O$ angles within  the $SiO_4$ tetrahedron are relatively uniform
ranging  from   $108.3^o$ to   $110.7^o$ with  a   $Si-O-Si$ angle  of
$144.0^o$ \cite{Jorgensen1978}. These  differences are important as it
results in   different   mechanisms being  responsible   for  the high
pressure   behaviour  of  $\alpha$-quartz   and   $\alpha$-quartz-like
$GeO_2$. For $\alpha$-quartz  the tetrahedra are relatively rigid  and
compression  of  the structure   occurs via  cooperative   rotation or
tilting  of    the  tetrahedra      around   the  shared      bridging
oxygens.    Conversely,  for    $\alpha$-quartz-like    $GeO_2$  while
compression does  occur via tilting of  the tetrahedra,  distortion of
each tetrahedron  via changes in  the  individual $O-Ge-O$ angles also
plays a large role  \cite{Jorgensen1978}. This behaviour is the reason
why germanate analogues of silicate phases are useful in high pressure
studies since they  undergo pressure induced phase transformations  at
much lower pressures than their silicate analogues. The transformation
of $\alpha$-quartz to the high  pressure rutile structure (Stishovite)
occurs  at  $10$    GPa   while the   equivalent transformation    for
$\alpha$-quartz-like $GeO_2$ to rutile-like  $GeO_2$ has been observed
to occur at much  lower pressures when  the sample is  heated: $\simeq
1.8$-$2.2~GPa$ at  $417 K$ \cite{Ault1996}.  \par  As noted above, the
stable room temperature $GeO_2$ phase is  the rutile $GeO_2$ polymorph
which transforms  to  the $\alpha$-quartz-like structure   at $1281~K$
(\cite{Laubengayer1932},      \cite{Yamanaka1992}     and   references
therein). The rutile $GeO_2$ polymorph has a structure similar to that
of stishovite \cite{Baur1971} and, like stishovite,  the 2 axial bonds
within the $GeO_6$ polyhedron  are   longer than the  $4$   equatorial
$Ge-O$   bonds: $1.902±0.001~Å$    and        $1.872±0.001~Å$,
respectively. Conversely, the  two independent $Ge-O$ distances in the
$\alpha$-quartz-like $GeO_2$   structure   are  similar  at  $1.737±
0.003~Å$ and $1.741± 0.002~Å$ \cite{Smith1964}.
\begin{figure}
\begin{center}
\includegraphics[width=0.45\linewidth]{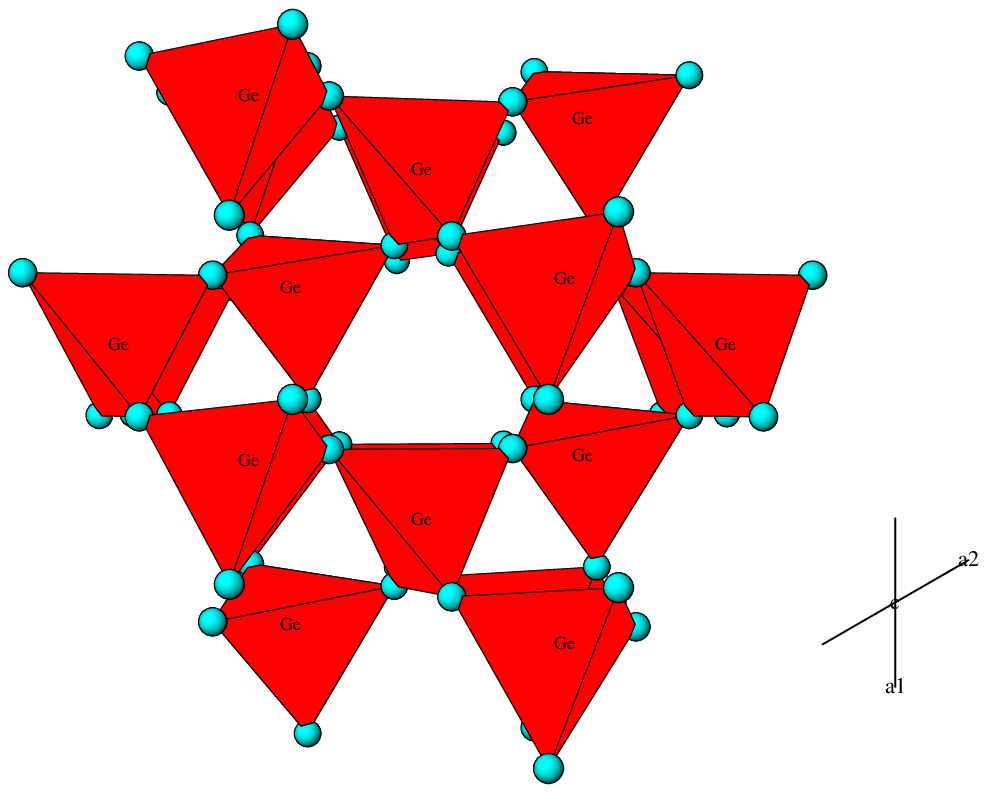}
\includegraphics[width=0.45\linewidth]{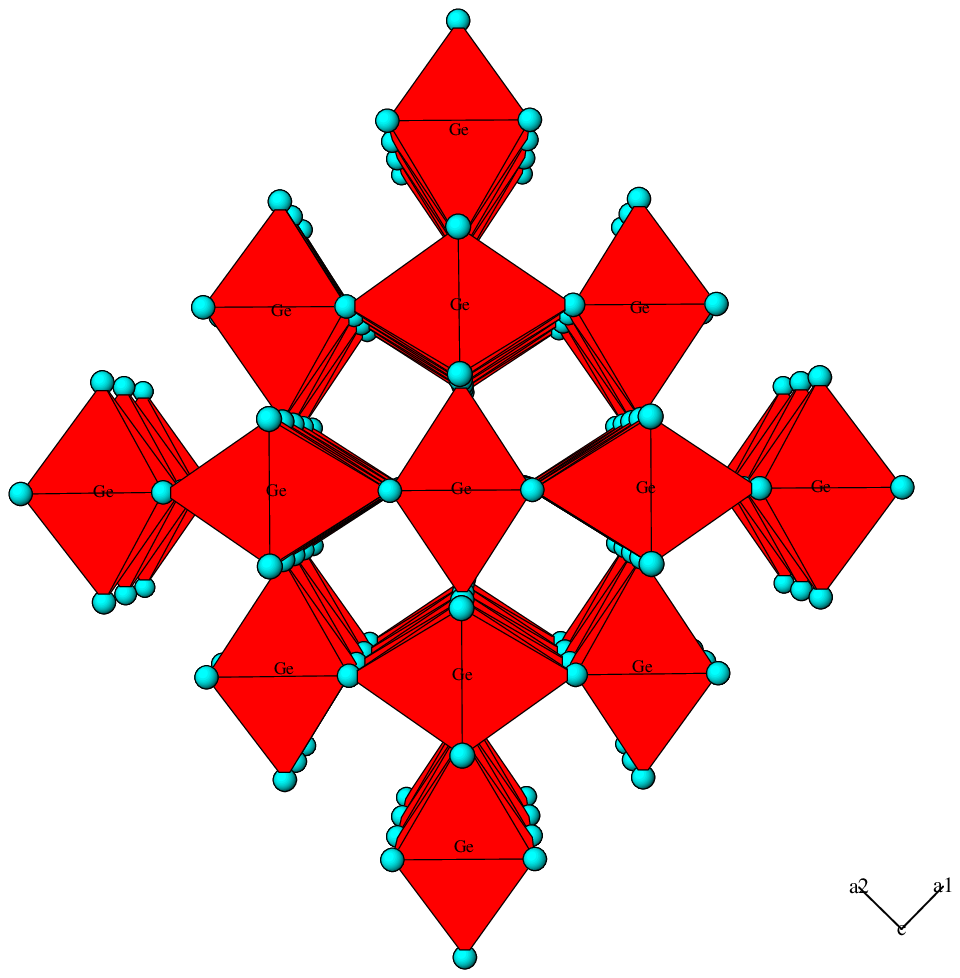}
\end{center}
\caption{Projection of the $\alpha$-quartz-like structure (left) and 
rutile-like structure (right) onto the (001) plane.}
\end{figure}

\subsection{High pressure and temperature behaviour}

A number of  studies have investigated  the high pressure behaviour of
the two $GeO_2$  polymorphs. Itié et al.  \cite{Itie1989} investigated
$\alpha$-quartz-like $GeO_2$ at ambient  temperature. They observed an
increase in the $Ge-O$    bond  length and $Ge$  coordination   number
consistent with the formation of the rutile-like $GeO_2$ phase between
$7-9~GPa$. However,   subsequent  studies   have suggested  that   the
transformation is  to an amorphous  phase rather than  the crystalline
rutile-like   $GeO_2$ polymorph \cite{Yamanaka1992},   \cite{Wolf1992},
\cite{Kawasaki1994},   \cite{Tsuchiya1998}.  Furthermore, it  has been
suggested  that the amorphization  step is  a  precursor to subsequent
transformation to the rutile polymorph \cite{Tsuchiya1998}.  More
recently,  Brazhkin et al.  \cite{Brazhkin2000a},\cite{Brazhkin2000b},
\cite{Brazhkin2003}  have  shown   that  with compression, $\alpha$-$GeO_2$
changes   via a martensitic transition  into  a crystalline monoclinic
($P2_1/c$) phase.  On the other hand, Haines et al. \cite{Haines2000a}
suggest that  there is    no   evidence  for  amorphization    of  the
crystal. Instead  a  poorly  crystalline monoclinic ($P2_1/c$)   phase
forms consisting  of edge sharing  chains of $GeO_6$ octahedra (Figure
2).  \par The monoclinic phase  is metastable up to $50~GPa$. However,
when combined with heating,  it transforms to  the rutile structure at
pressures up  to  $22~GPa$ and   above $43~GPa$  forms  a  mixture  of
$CsCl_2$-type  and   $Fe_2N$-type (or   $\alpha$-$PbO_2$, see  later)  high
pressure phases
\cite{Brazhkin2000a},\cite{Brazhkin2000b},\cite{Haines2000a}.   This
monoclinic phase      was    also   reported    by   Prakapenka     et
al. \cite{Prakapenka2003} between  $7-52~GPa$ at  room temperature but
with  laser heating  it  transforms  to an orthorhombic  $CaCl_2$-type
structure   above $36.4~GPa$   and   an $\alpha-PbO_2$-type  structure   at
$41~GPa$.

\begin{figure}
\begin{center}
\includegraphics[width=0.7\linewidth]{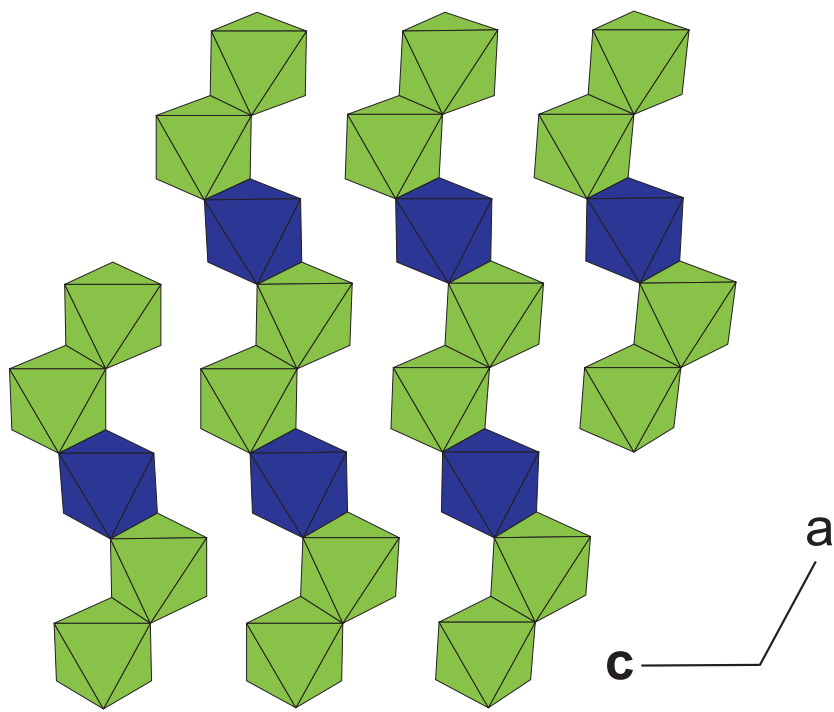}
\end{center}
\caption{Polyhedral representation of the ($3× 2$)-kinked $P2_1/c$ structure of 
$GeO_2$    determined     by  Haines   et      al. \cite{Haines2000a}.
Crystallographic  axes are  not   to  scale and   are  merely  to show
orientation  of the structure. Green   (light) octahedra are the fully
occupied  Ge  positions while the  blue   (dark) octahedra are  the Ge
octahedra which exhibit partial occupancy.}
\end{figure}

\par 

Haines   et   al.     \cite{Haines2000b}  have   also    observed
transformation  of the rutile-like  $GeO_2$  phase to the orthorhombic
$CaCl_2$-type structure  above $25~GPa$  at ambient  temperature while
Ono et al. \cite{Ono2002} observed the transition at high pressure and
temperature.     {\em Ab    initio}    calculations   by  Lodziana  et
al.  \cite{Lodziana2001}  suggested  that  rutile-type  $GeO_2$ should
transform to $\alpha$-$PbO_2$-type  (above $\simeq 36~GPa$) and pyrite
(Pa)   type   (above $\simeq65.5~GPa$)  structures    and these   were
subsequently observed  by Ono  et al. \cite{Ono2003a}, \cite{Ono2003b}
around $44$  and  $90~GPa$,  respectively. A $Fe_2N$-type   (or defect
$Ni-As$) phase at pressures larger than  $25~GPa$ has been observed by
Liu et  al. \cite{Liu1978} and  Haines et al. \cite{Haines2000a}. This
type of structure is similar to an $\alpha$-$PbO_2$-type structure but
with the $Ge$ sites disordered and has, more recently, been explicitly
identified  by  Ono       et  al.   \cite{Ono2003a}  as   being    the
$\alpha$-$PbO_2$-type  structure. However,   it should be  noted  that
Prakapenka  et  al.  \cite{Prakapenka2004}  observe  the  defect $NiAs$
structure when amorphous $GeO_2$ is heated to $1000-1300~K$ at $6~GPa$
(see later).

\par

Structural refinements of the crystalline phases have been obtained by
Shiraki et al. \cite{Shiraki2003} and  a phase diagram for crystalline
$GeO_2$ is given in Figure 3.  In addition, another orthorhombic phase
has been suggested to occur at $\simeq28~GPa$ and $1273~K$ by Ming and
Manghnani \cite{Ming1983}. They concluded that  this phase was not the
$\alpha$-$PbO_2$-type structure   but   it  has  not  been    observed
subsequently. The phase transformation sequence of rutile-like $GeO_2$
$\to$   ($25~GPa$)    $CaCl_2$-type   $\to$ ($44~GPa$)
$\to$     $\alpha$-$PbO_2$-type  $\to$   ($70-90~GPa$)
pyrite-type structures is  consistent with the high-pressure behaviour
of other  group-IV  element  dioxides such  as  $PbO_2$, $SnO_2$,  and
$SiO_2$     (cf.,       \cite{Teter1998},         \cite{Lodziana2001},
\cite{Prakapenka2003}, \cite{Prakapenka2004}).

\begin{figure}
\begin{center}
\includegraphics[width=0.8\linewidth]{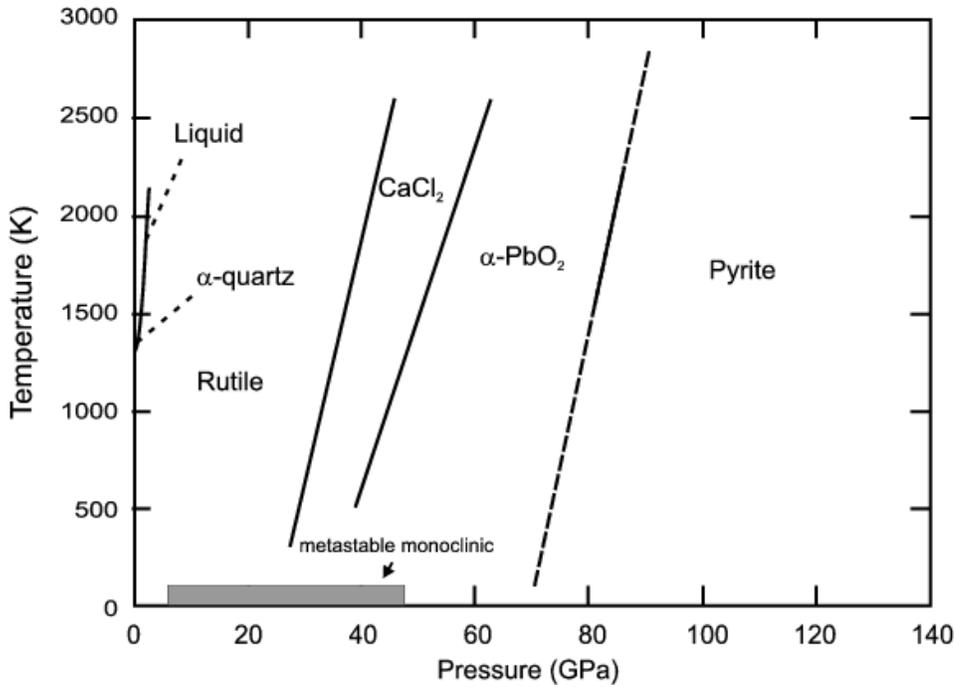}
\end{center}
\caption{Phase diagram of crystalline $GeO_2$ (after \cite{Ono2003a}).}
\end{figure}

Of interest  is  the   way   in which   the  $\alpha$-quartz-like  and
rutile-like $GeO_2$   structures respond  to increasing   pressure. As
noted  above, Jorgensen   \cite{Jorgensen1978}  observed  that  in the
$\alpha$-quartz-like  polymorph compression   up to $2.5~GPa$   occurs
predominantly {\em via} changes  in the individual $O-Ge-O$ angles and
that tilting  of   tetrahedra   was  secondary. Yamanaka   and   Ogata
\cite{Yamanaka1991} carried out a series  of structural refinements on
the $\alpha$-quartz-like $GeO_2$ polymorph up  to $4.48~GPa$ and found
that the $GeO_4$ tetrahedra are relatively rigid with little change in
the   $Ge-O$  bond length,  consistent  with  the   study  of Itié  et
al. \cite{Itie1989}. Yamanaka and Ogata \cite{Yamanaka1991} found that
the  dominant mechanism responsible  for the observed pressure induced
unit-cell volume change in the  structure involved  a decrease in  the
$Ge-O-Ge$ angle from   $130^o$ to  $125^o$. Conversely, Glinneman   et
al.     \cite{Glinnemann1992}  found  that   tetrahedral   tilting  was
responsible for the   $11\%$  volume  change  of  $\alpha$-quartz-like
$GeO_2$ up to $5.57~GPa$.

\par 

The phase  transformation of rutile-like  $GeO_2$ to the $CaCl_2$-type
structure occurs {\em via} compression  of the  axial $Ge-O$ bonds  of
the  octahedron.  The axial   bonds  are  elongated  relative  to  the
equatorial bonds  (see   above). With  increasing pressure  there   is
increased compression of the    axial  relative to   equatorial  bonds
\cite{Haines2000b}  and the transformation   at $25~GPa$ occurs during
flattening           of       the   octahedra      \cite{Haines2000b},
\cite{Shiraki2003}.  With transformation  to the $\alpha$-$PbO_2$-type
structure, the $GeO_6$  octahedron becomes  further deformed with  the
$Ge$ atom displaced from the center  of the octahedron  and $2$ of the
six $Ge-O$ bonds  becoming elongated \cite{Shiraki2003},  as suggested
by       the        numerical      results     of       Lodziana    et
al.  \cite{Lodziana2001}. Transformation to the pyrite-type structure,
however, results in  $GeO_6$ octahedra that  are symmetrical with $Ge$
in the centre.

\par

As noted above, the $\alpha$-quartz-like  polymorph is the stable high
temperature phase  and  rutile-type  $GeO_2$  will transform   to this
polymorph above  $1320~K$;  the  transformation temperature  being the
highest of any of the quartz-like  analogues. The high temperature (up
to   $1344~K$) behaviour of this   polymorph  has been investigated by
Haines  et al. \cite{Haines2002} who   found that the intertetrahedral
bridging angle ($Ge-O-Ge$) and tilt angles exhibit thermal stabilities
that are amongst the  highest observed for quartz-type analogues. With
increasing temperature,    expansion  of  the   unit  cell  is  highly
anisotropic with expansion along {\bf a}  being $5$ times greater than along
{\bf c}     \cite{Haines2002}.    However, the  $\alpha$-quartz-like $GeO_2$
polymorph is   metastable   at low   temperatures (\cite{Balitsky2001}
provide  a number of   methods  for growing   the $\alpha$-quartz-like
$GeO_2$ polymorph) but does  undergo transformation to the rutile-type
polymorph at around $1000~K$ although the reaction proceeds slowly due
to the kinetics involved (cf. \cite{Madon1991}). Finally, it should be
noted that a cristobalite-like polymorph for $GeO_2$ has been observed
after long-time heating of $GeO_2$ glass to $873~K$ \cite{Bohm1968} or
by         dehydration  of       ammonium         hydrogen   germanate
($(NH_4)_3HGe_7O_{16.4}H_2O$)  between  $853-873~K$ \cite{Hauser1970},
however, this polymorph  has not been observed  in {\em in-situ}  high
pressure  and    temperature     studies.      In      addition,   the
$\alpha$-quartz-like polymorph at $1322~K$ mentioned by Leadbetter and
Wright \cite{Leadbetter1972} and Desa  et al. \cite{Desa1988} based on
the work of  Laubengayer and Morton \cite{Laubengayer1932}  and Sarver
and Hummel \cite{Sarver1960} has also not been observed.

\section{$GeO_2$ glass structure}

\subsection{Neutron and X-ray diffraction}

Neutron and  X-ray diffraction    data  are complementary   tools  for
inferring  structural  information since  the  chemical sensitivity is
different for the two techniques; $Ge-O$  and $Ge-Ge$ pairs are better
resolved with X-ray and $Ge-O$ and $O-O$ with neutron.

\par

$GeO_2$ glass structure  has  been studied using X-ray  diffraction in
the pioneering work  of  Warren \cite{Warren1934a}, \cite{Warren1934b}
and Zarzycki \cite{Zarzycki1956},   \cite{Zarzycki1957}. It was  found
that the $Ge$  atoms are arranged in  basic tetrahedral units  such as
those    found    in  the   trigonal     $\alpha$-quartz-like  $GeO_2$
polymorph. X-ray diffraction  data with  higher real space  resolution
($Q_{max} = 17~Å^{-1}$)         confirmed     these         findings
\cite{Leadbetter1972} and   determined the  first  $Ge-O$ and  $Ge-Ge$
distances at   $1.74~Å$  and  $3.18~Å$,  respectively,  giving  an
intertetrahedral   angle  of    $\simeq~133^o$.   The  first   neutron
diffraction  experiment ($Q_{max} = 18~Å^{-1}$)  on vitreous  $GeO_2$
shows two strong peaks at $1.72~Å$ and $2.85~Å$ ascribed to $Ge-O$
and $O-O$ correlations,  which  is consistent with $GeO_4$  tetrahedra
\cite{Lorch1969}. The $Ge-Ge$ peak, initially determined at $3.45~Å$
\cite{Lorch1969},   \cite{Ferguson1970},   was  resolved  in   a high
resolution       neutron        diffraction              investigation
($Q_{max} = 35.5~Å^{-1}$) at $3.21~Å$, which  is slightly higher than
the  $Ge-Ge$ distance   determined by X-ray  diffraction  due  to  the
overlapping of    $Ge-O$    and  $O-O$   pairs    \cite{Sinclair1974},
\cite{Sinclair1977}.   A  recent    neutron   and   X-ray  diffraction
investigation   \cite{Desa1988}  has    shown    that  the    $O-Ge-O$
intratetrahedral  angle is more distorted  in vitreous $GeO_2$ than in
vitreous   $SiO_2$, with a distribution  likely  comparable to that of
$GeO_2$  $\alpha$-quartz ($106.3-113.1^o$). This is due  to the larger radius
of  Ge than Si, allowing more  accessible positions for O atoms around
Ge atoms.  The mean Ge-O-Ge  intertetrahedral angle was estimated from
the   Ge-O and Ge-Ge   distances  to  be  $130.1^o$   with a range  of
$121-147^o$. This  mean value was  confirmed  at $133±  8.3^o$ using
high-energy  X-ray  diffraction \cite{Neuefeind1996}.  This bond angle
and  its   distribution are   lower than in     the case  of  vitreous
silica.  The smaller Ge-O-Ge angle probably  results from the presence
of increased    numbers of 3-membered rings   in   the $GeO_2$ network
relative to vitreous $SiO_2$ (cf. \cite{Desa1988} and see later) since
such   planar    rings   have   a    Ge-O-Ge   angle    of   $130.5^o$
\cite{Galeener1982},\cite{Barrio1993}.  The   values    for  the  main
interatomic  distances,   coordination   numbers and  intertetrahedral
angles found in these  studies are reported in  Table I. The structure
of $GeO_2$ can thus be  viewed as continuous  random network of corner
sharing tetrahedral, as in silica,  but with greater distortion of the
tetrahedra and larger amounts of three-membered rings.

\par 

The diffraction data  (Fig. 4) of  $GeO_2$  are  composed   of  three  partial
functions, Ge-Ge, Ge-O  and O-O.  The  first attempt to separate   the
three components was carried out using X-ray anomalous diffraction and
neutron  diffraction  \cite{Bondot1974a}, \cite{Bondot1974b}.  The Ge-O,
O-O    and  Ge-Ge distances   are    found  at   $1.73$,  $2.85$,  and
$3.17± 0.04 Å$,    respectively,   and    the   average    Ge-O-Ge
intertetrahedral bond  angle   is estimated  to   have values  between
$129^o$  and $139^o$.   Recently,    by combining neutron   and  X-ray
diffraction,  together   with X-ray   anomalous  scattering, the three
partial  functions  were   fully   separated  up   to  $Q = 9~Å^{-1}$
\cite{Waseda1990},\cite{Price1998}, though   problems  exist  due   to
different instrumental  resolution functions that appear especially at
low $Q$ values, and the  necessity to improve the anomalous scattering
terms \cite{Barnes1999}. The structure  factors are dominated by peaks
occurring at  $1.54$, $2.6$ and  $\simeq  4.5Å^{-1}$. The first  feature at
$1.54~Å^{-1}$  (usually  called  first  sharp  diffraction peak, FSDP)
corresponds  to intermediate range  ordering  and is stronger in X-ray
than in neutron data \cite{Price1997}.  The FSDP is associated with  a
positive peak in $S_{GeGe}$ and $S_{GeO}$ and  a shallow negative peak
in $S_{OO}$,  indicating that cation  correlations dominate the medium
range order (Figure 5,  \cite{Price1998}). The peak at $2.6~Å^{-1}$  is
strong  and  positive  in $S_{GeGe}$    and $S_{OO}$, and  strong  and
negative in $S_{GeO}$, and has been attributed to chemical short range
order. The peak at $4.4~Å^{-1}$ occurs predominantly in $S_{GeGe}$ and
is due to topological  short  range order.   The latter conclusion  is
usually extracted  from Bhatia-Thornton structure factors \cite{Bhatia1970}
that
show the correlations between number density and concentration fluctuations
\cite{salmon2005}. On this basis, the chemical and topological ordering in 
$GeO_2$  can be rationalized in   terms  of an  interplay between  the
relative importance of  two  lengthscales  that  exist in  the   glass
\cite{Salmon2006}.
\begin{figure}
\begin{center}
\includegraphics[width=0.8\linewidth]{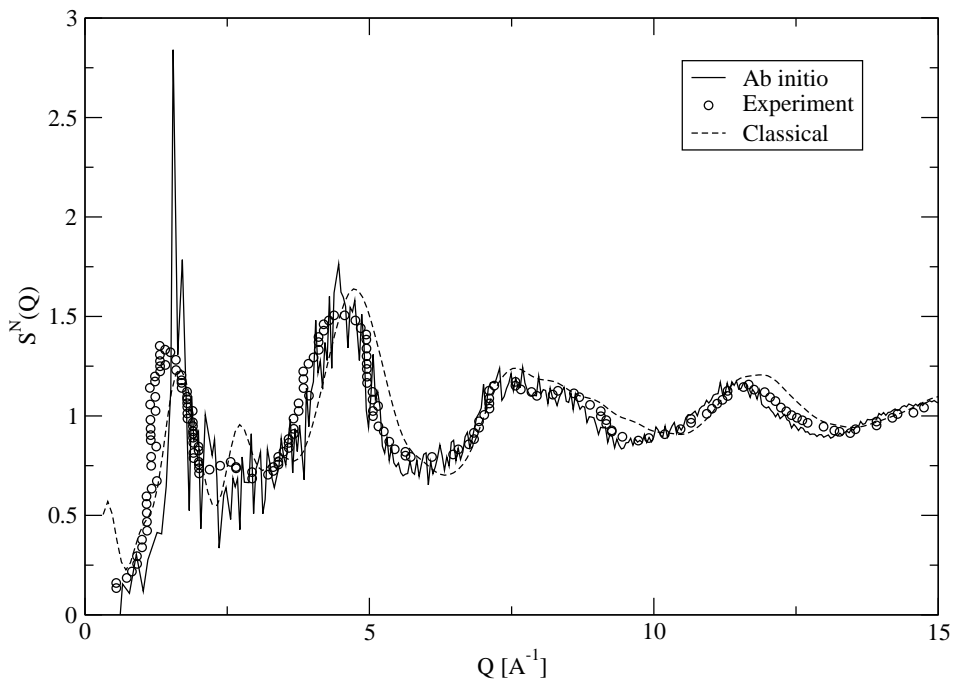}
\end{center}
\caption{Measured total structure factor (circles, \cite{Stone2001}) together with calculated $S(Q)$
from {\em ab initio} (solid line, \cite{Giacomazzi2005}) and Classical
Molecular   Dynamics   (broken line,   \cite{Micoulaut2006}).}
\end{figure}

\begin{figure}
\begin{center}
\includegraphics[width=0.45\linewidth]{sqn.eps}
\includegraphics[width=0.45\linewidth,height=0.875\linewidth]{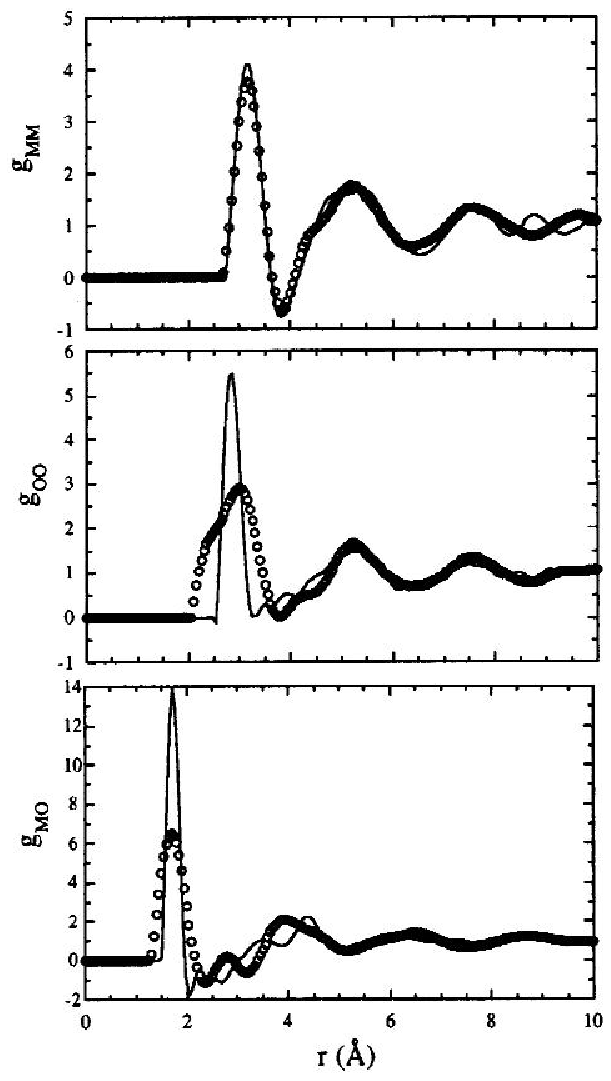}
\end{center}
\caption{Left from top to bottom: Measured
(circles,    \cite{Price1998})   and   calculated   (solid       line,
\cite{Micoulaut2004a})   partial  structure  factors,   $S_{GeGe}(Q)$,
$S_{GeO}(Q)$  and $S_{OO}(Q)$.  Right : Partial
correlation functions,  $g_{MM}(r)$, $g_{MO}(r)$ and $g_{OO}(r)$ (from
top to bottom),  for the three atom pairs  in vitreous $GeO_2$ at room
temperature (M  =   Ge,  points),  together  with  the   corresponding
functions  from  rescaled  Molecular Dynamics simulation  of
vitreous $SiO_2$ (M = Si, lines) (After \cite{Price1998}).}
\end{figure}

\par

There  have been considerable efforts    to compare diffraction   data
obtained  on $GeO_2$ glass with equivalent  calculations  based on the
$GeO_2$ crystalline polymorphs, with divergent results. Leadbetter and
Wright  \cite{Leadbetter1972} concluded   that the  intermediate range
order in the glass  closely resembles a quasi-crystalline  model based
on the $\alpha$-quartz-like   $GeO_2$  structure  with  a  correlation
length  of $10.5~Å$ but  discrepancies appear beyond $4~Å$. Bondot
\cite{Bondot1974b}  obtained good agreement between  the glass and the
$\alpha$-  and  $\beta$-quartz  $GeO_2$ polymorphs,  which led  to the
conclusion  that  the  glass   contains  six-membered  rings.  On  the
contrary, Konnert  et al.,  \cite{Konnert1973} concluded that vitreous
germania, like vitreous silica,  possesses the same short  range order
as that found in the tridymite $SiO_2$ polymorph. The vitreous $GeO_2$
structure  could thus be   described  as randomly  oriented,  slightly
distorted  tridymite-like regions having   dimensions ranging up to at
least   $20Å$  \cite{Konnert1973}. However, these   regions are not
crystallographically ordered  (i.e. not microcystals) but have similar
bonding topology in the glass and in tridymite. In a more recent study
\cite{Desa1988}, it  was  shown that, though  similarities exists with
crystalline   $\alpha$-quartz     and    $\alpha$-cristobalite $GeO_2$
polymorphs, diffraction data   are not  consistent with  large  volume
fractions  of   quasi crystalline-like  regions,  due to  an important
distribution of torsion angles.

\par

\begin{table}
\begin{center}
\begin{tabular}{cccccc}\hline

Pair ij&$R (Å)$&$N$&$\sigma(Å)$&Method&Ref\\ \hline\hline
Ge-O&$1.733± 0.001$&$3.99± 0.1$&$0.042±0.001$&ND&1 \\
 &$1.744± 0.05$&$4.0± 0.2$&$0.11± 0.01$&ND&2 \\
 &$1.73±0.03$& & &ND+AXS&3 \\
 &$1.74± 0.01$&$3.7± 0.2$& &ND&4 \\
 &$1.75$& & &ND+AXS&5 \\
 &$1.73$& & &HEXRD&6 \\
 &$1.739±0.005$&$3.9± 0.1$& &D&7 \\
 &$1.73$& & &AXS&8 \\
 &$1.74$& & &XRD&9 \\ \hline
O-O&$2.822± 0.002$&$6.0*$&$0.100± 0.002$&ND&1 \\
 &$2.84± 0.01$&$6.0±0.3$&$0.26± 0.03$&ND&2 \\
 &$2.83± 0.05$& & &ND+AXS&3 \\
 &$2.84± 0.02$&$5.5± 0.5$& &ND&4 \\
 &$2.82$& & &ND+AXS&5 \\
 &$2.838$&$6.0*$&$0.109$&ND&7 \\ \hline
Ge-Ge&$3.155± 0.01$&$4.0±0.3$&$0.26±0.03$&ND&2 \\
 &$3.16±0.03$& & &ND+AXS&3 \\
 &$3.18±0.05$& & &ND&4 \\
 &$3.18$& & &ND+AXS&5 \\
 &$3.17$& & &HEXRD&6 \\
 &$3.185$&4.0*&$0.163$&ND&7 \\
 &$3.17$& & &AXS&8 \\
 &$3.18$& & &XRD&9 \\ \hline
Angle&Ge-O-Ge& & & \\
 &$132± 5^o$& & &ND+AXS&3 \\
 &$133± 8.3^o$& & &HEXRD&6 \\
 &$130.1^o$& & &ND+XRD&7 \\
 &$133^o$& & &XRD&9 \\
\end{tabular}
\end{center}
\caption{Interatomic distances ($R$), Coordination numbers ($N$), standard 
deviations ($\sigma$) and Ge-O-Ge intertetrahedral angle determined by diffraction methods.
1) $Q_{max} = 50~Å^{-1}$, \cite{Stone2001}; 2) $Q_{max} = 50~Å^{-1}$, \cite{Hoppe2000}; 
3) $Q_{max} = 9~Å^{-1}$, \cite{Price1998}; 4) \cite{Price1997}; 5) $1~Å^{-1}\leq Q \leq 
10~Å^{-1}$, \cite{Waseda1990}; 6) $0.6~Å^{-1}\leq Q\leq 33.5~Å^{-1}$, 
\cite{Neuefeind1996}; 7) $0.22~Å^{-1}\leq Q\leq 23.6~Å^{-1}$, \cite{Desa1988}; 
8) \cite{Bondot1974b}; 9) $0.8~Å^{-1}\leq Q\leq 17~Å^{-1}$, \cite{Leadbetter1972}.
ND = Neutron diffraction; AXS = anomalous X-ray scattering; HEXRD = high energy X-ray 
ray diffraction; XRD = X-ray diffraction. * Fixed values.}
\end{table}

\subsection{Neutron and X-ray diffraction at high pressure and temperature}

Due to the technical  difficulties associated with performing {\em  in
situ}  diffraction  experiments,  pressure   effects have been  mainly
studied on pressure-released glasses, in which permanent densification
is  observed.  Permanently densified  glasses  (up  to  $18~GPa$) were
studied  by X-ray diffraction   in the low   Q-region (FSDP)  which is
sensitive to  medium range order \cite{Sugai1996a}.   A shift to higher
$Q$ and an increase in width of the FSDP is  observed above $6~GPa$, a
pressure corresponding  to  the   threshold for coordination   changes
observed in {\em in  situ} experiments (see below).  However, comments
on this study  pointed out that changes  in the diffraction peaks  may
not    necessarily  be  associated    with    a coordination    change
\cite{Hemley1997}.

\par

$GeO_2$ glasses densified up  to $6~GPa$ at $673~K$  (densification of
$16\%$) were   investigated by neutron  diffraction  \cite{Stone2001},
while a  glass  densified at  $10~GPa$ and $300~K$  (densification  of
$11\%$)  was studied by  neutron   and  X-ray diffraction (Figure   6)
\cite{Sampath2003}.    No   evidence     of  six-coordinated Ge    was
observed. The $GeO_4$  tetrahedra are distorted, with $Ge-O$ distances
increasing  by $0.005± 0.001~Å$   and $O-O$ and  $Ge-Ge$ distances
decreasing   by   $0.023 ± 0.002~Å$     and  $0.019± 0.002~Å$,
respectively  \cite{Sampath2003}. The  main  change is  a shift of the
$Ge-Ge$ peak (at $\simeq 3.1~Å$) to lower $r$ values with increasing
pressure compaction \cite{Stone2001}.   This indicates  a reduction in
the mean $Ge-O-Ge$ bond angle    with increasing density.   Noticeable
changes  are  seen for the  FSDP  in  the neutron and  X-ray structure
factors: the FSDP shifts towards  higher $Q$, broadens and become less
intense on  densification. This indicates  a reduction of  the network
connectivity.    By  combining neutron and    X-ray  diffraction up to
$Q=30~Å^{-1}$, it was  shown  \cite{Sampath2003} that  variations of
the  FSDP are  mostly  associated with $O-O$  correlations rather than
$Ge-Ge$ ones. This is attributed to a decrease  in the average size of
the network cages (these can be considered as holes in the structures,
formed for instance, by  the ring structures), yielding better packing
of the $GeO_4$ tetrahedra.

\par

{\em In situ}  measurements have recently been  obtained (Figure 6) by
both neutron  (up to $5~GPa$) and X-ray  (up to  $15~GPa$) diffraction
\cite{Guthrie2004}.  The  FSDP   decreases  and  almost  vanishes with
increasing pressure in  neutron measurements while it gradually shifts
to higher $Q$ in X-ray data. This is interpreted as a breakdown of the
intermediate range order upon  compaction  of the tetrahedral  network
associated with  changes  in the  oxygen  correlations. In  the  X-ray
correlation functions, a reduction of  the $Ge-O$ distance is observed
below $6~GPa$ while it increases at further pressure, corresponding to
$GeO_4$  tetrahedra being converted  to  $GeO_6$  octahedra. Based  on
molecular  dynamics simulations,  it was argued  that stable five-fold
units  are   present  in  the  transition  region,  indicating   a new
intermediate form of the   glass. The structure  of the  high pressure
glass is  based on  edge- and  corner-shared  octahedra, which is  not
retained upon decompression.

\par

$GeO_2$ in the liquid state has been investigated by X-ray diffraction
\cite{Zarzycki1957},   \cite{Kamiya1986}.   The  $Ge-O$  distance   is
unchanged in  agreement with a small  thermal expansion  of the $Ge-O$
bond  similar  to that for  $Si-O$  bonds.  The $GeO_4$ tetrahedra are
preserved in the  $GeO_2$ melt but $Ge-Ge$  distances are shifted from
$3.16~Å$  at room temperature to $3.25~Å$   at $1100^oc$, which is
interpreted as a widening of the $Ge-O-Ge$ bond angle.

\begin{figure}
\begin{center}
\includegraphics[width=0.8\linewidth]{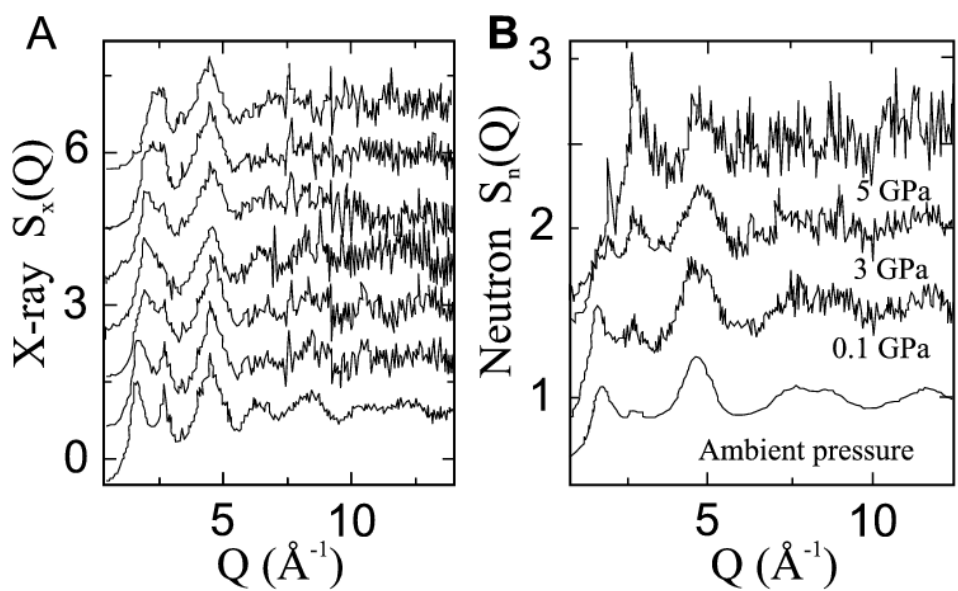}
\end{center}
\caption{{\em In situ} structure factors (after \cite{Guthrie2004} for 
(A) X-ray diffraction at $0$, $3$, $5$, $6$, $7$, $10$, and $15~GPa$ (bottom to top) 
and (B) neutron diffraction up to $5~GPa$, with ambient-pressure data 
from Sampath et al. \cite{Sampath2003}.} 
\end{figure}

\subsection{Raman Spectroscopy}

\subsubsection{$GeO_2$ polymorphs}

The Raman spectra of the crystalline polymorphs  of $GeO_2$ (Figure 7)
were first reported by Scott \cite{Scott1970}. The rutile-like $GeO_2$
spectrum exhibits three strong  bands in the  $150-1200~cm^{-1}$ range
at $173$,  $701$ and $873~cm^{-1}$. The band  at $701~cm^{-1}$  is the
$A_{1g}$  mode while the $873~cm^{-1}$ band  is the $B_{2g}$ mode. The
$B_{1g}$ mode   is  at  $173~cm^{-1}$.  The $E_g$    mode  observed at
$680~cm^{-1}$ by  Scott   \cite{Scott1970} is   not observed  in   the
spectrum shown in figure 7. Alpha-quartz-like $GeO_2$  has a number of
additional bands including four symmetric modes  of $A_1$ symmetry and
8  doubly degenerate modes of  $E$ symmetry  all split into transverse
optic (TO)  and longitudinal optic   modes (LO) \cite{Scott1970}.  The
$\alpha$-quartz-like  $GeO_2$ spectrum of   figure 7 is comparable  to
that first  obtained by Scott  \cite{Scott1970}. Bands can be assigned
following Scott \cite{Scott1970}  and Dultz et al. \cite{Dultz1975} as
$A_1$ modes  at $263$, $330$,  $444$, and $881~cm^{-1}$;  $E$ modes at
$123$ (TO+LO), $166$  (TO+LO),  $212$ (TO),  $330$ (TO),  $516$  (LO),
$593$ (LO), $860$ (TO), $960$  (TO), and $973~cm^{-1}$ (LO). $E$ modes
at $372$  (LO), $385$ (TO), $492$ (TO),  $583$  (TO) and $949~cm^{-1}$
(LO) are too weak to be observed in the spectrum or are unlabelled for
clarity.

\begin{figure}
\begin{center}
\includegraphics[width=0.7\linewidth]{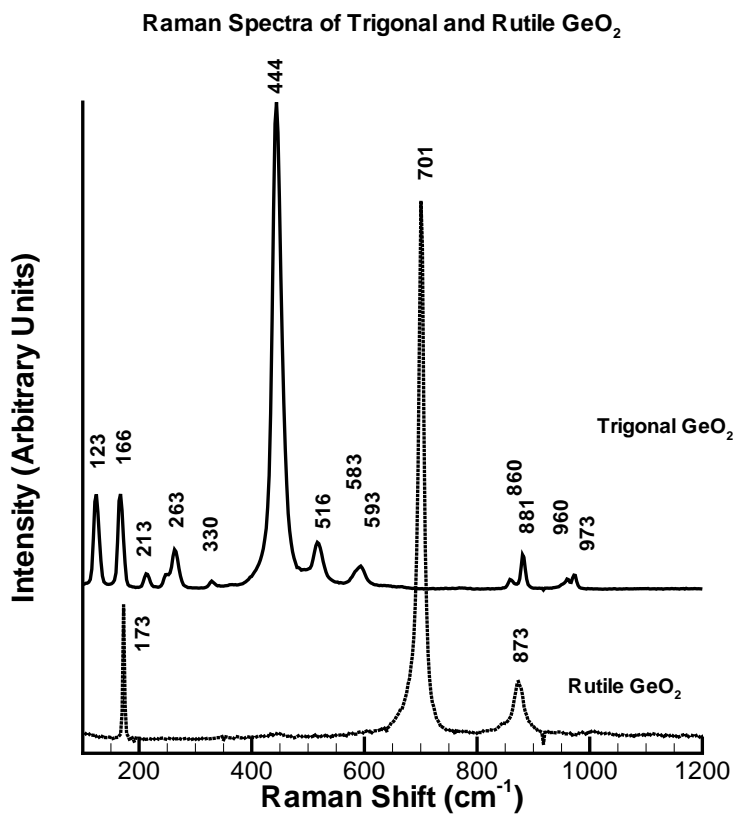}
\end{center}

\caption{Unpolarized Raman spectra of the trigonal and rutile $GeO_2$ polymorphs. Bands 
are comparable to those of Scott \cite{Scott1970} and Madon et al. \cite{Madon1991}. 
For clarity not all bands are labelled.}
\end{figure}

\par

The effects of increasing pressure  and temperature on the vibrational
spectra of the  $GeO_2$  polymorphs has  been investigated   by Sharma
\cite{Sharma1989}, Madon et  al. \cite{Madon1991} and Mernagh  and Liu
\cite{Mernagh1997}. With increasing  pressure the Raman  bands for the
$\alpha$-quartz-like polymorph shift  to higher wavenumber and  behave
in  a similar  manner as   the  IR modes   \cite{Madon1991}. The  mean
frequency  shift     is $\simeq   1~cm^{-1}/0.1~GPa$  for   bands   in
$400-600~cm^{-1}$ region, $0.3~cm^{-1}/0.1~GPa$ for  the bands  in the
$100-330~cm^{-1}$ region and the bands in the $850-970~cm^{-1}$ region
do   not shift   at all   up    to $4~GPa$  except  for   the  band at
$961~cm^{-1}$.   The  rutile Raman  bands  behave slightly differently
\cite{Sharma1989}  with the band  at  $\simeq 173~cm^{-1}$ shifting to
lower wavenumbers and the other two  bands to higher wavenumbers. With
increasing temperature, the  Raman bands of  the rutile-like polymorph
transform  to the  $\alpha$-quartz-like  spectrum at  $\simeq  1313~K$
\cite{Madon1991} while the  Raman  bands  of  the $\alpha$-quartz-like
polymorph  show  a nonlinear    shift with  increasing   T. Madon   et
al.  \cite{Madon1991} observed a  shift  of $-0.01~cm^{-1}/K$  for the
bands in the low  frequency region and  $-0.024~cm^{-1}/K$ in the mid-
and high-frequency regions. For  the rutile polymorph, the Raman bands
above  $600~cm^{-1}$  exhibit nonlinear  shifts  to lower  wavenumbers
whereas the $173~cm^{-1}$ band  exhibits a shift to higher wavenumbers
with   increasing T \cite{Mernagh1997}.  In   addition Mernagh and Liu
\cite{Mernagh1997} detect (by deconvolution) splitting of the $A_{1g}$
mode ($701~cm^{-1}$)  of the  rutile-like  polymorph with  a new  band
observed at $684~cm^{-1}$.
        
\subsubsection{$GeO_2$ glass and liquid}

The first Raman spectrum of   $GeO_2$ glass was described by  Bobovich
and   Tolub    \cite{Bobovich1958}     and     Obikhov-Denisov      et
al. \cite{Obikhov1960}. A Raman spectrum for $GeO_2$ glass is shown in
figure 8a. The Raman band assignments  for $GeO_2$ glass are similar to
those    of $SiO_2$  glass  but  are    shifted to lower   frequencies
(wavenumbers, $cm^{-1}$) because  of the larger  mass of $Ge$ relative
to $Si$. Currently accepted band  assignments for $GeO_2$ are given in
Table  2  and extensive discussion  of   Raman assignments and earlier
literature can be found in \cite{Henderson1985}, \cite{Henderson1991}.

\begin{figure}
\begin{center}a
\includegraphics[width=0.4\linewidth]{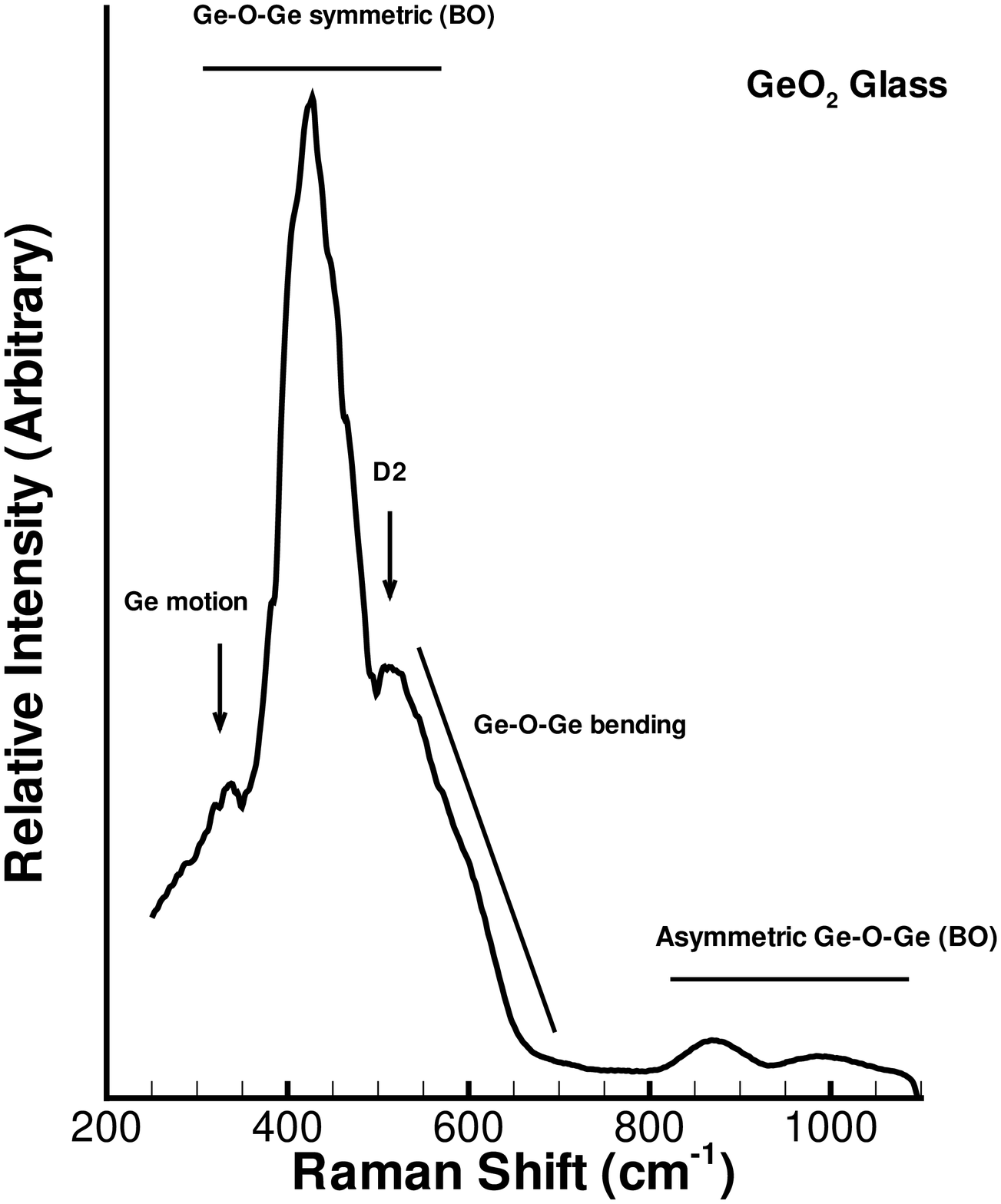}b
\includegraphics[width=0.4\linewidth]{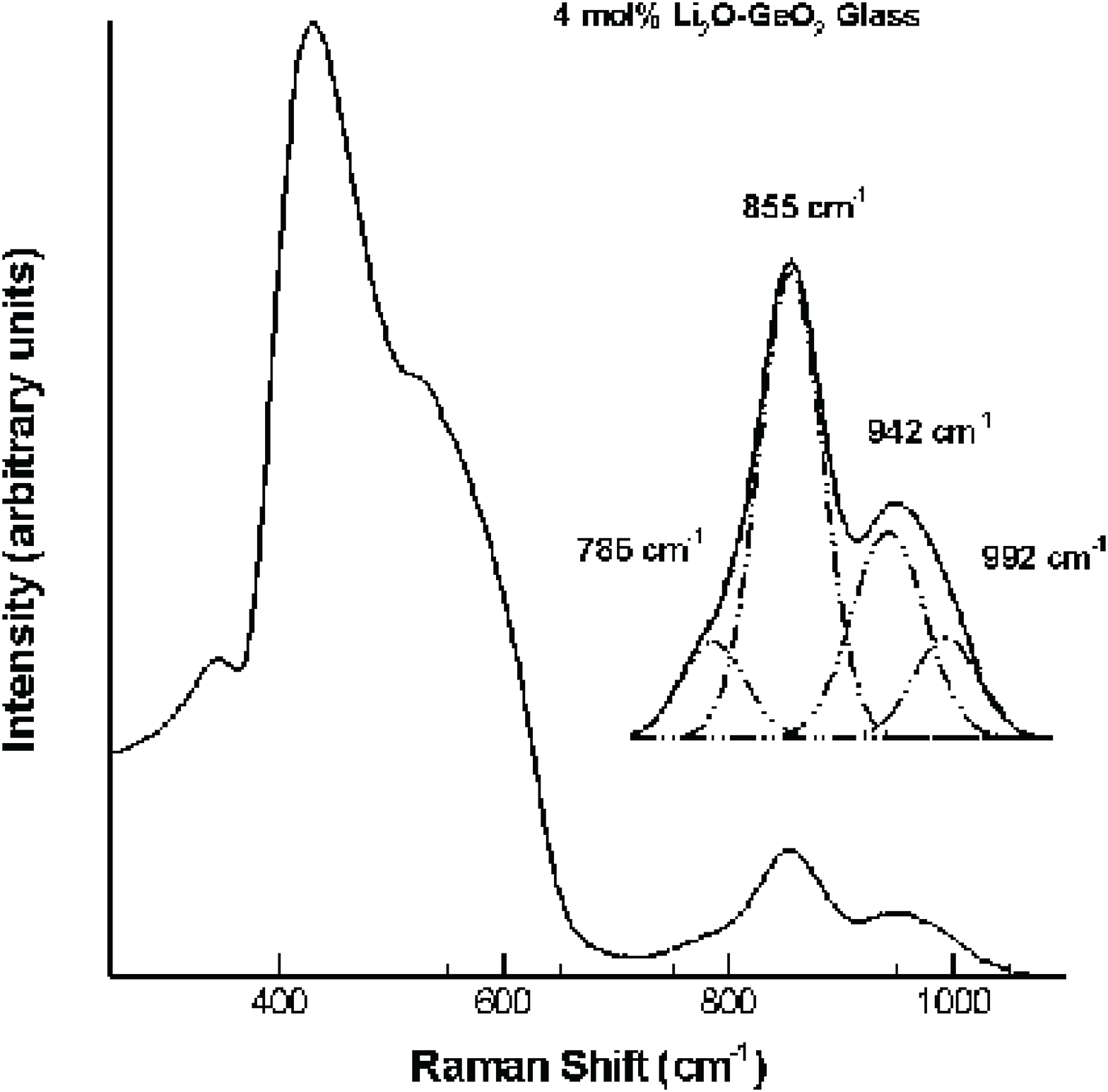}c
\includegraphics[width=0.4\linewidth]{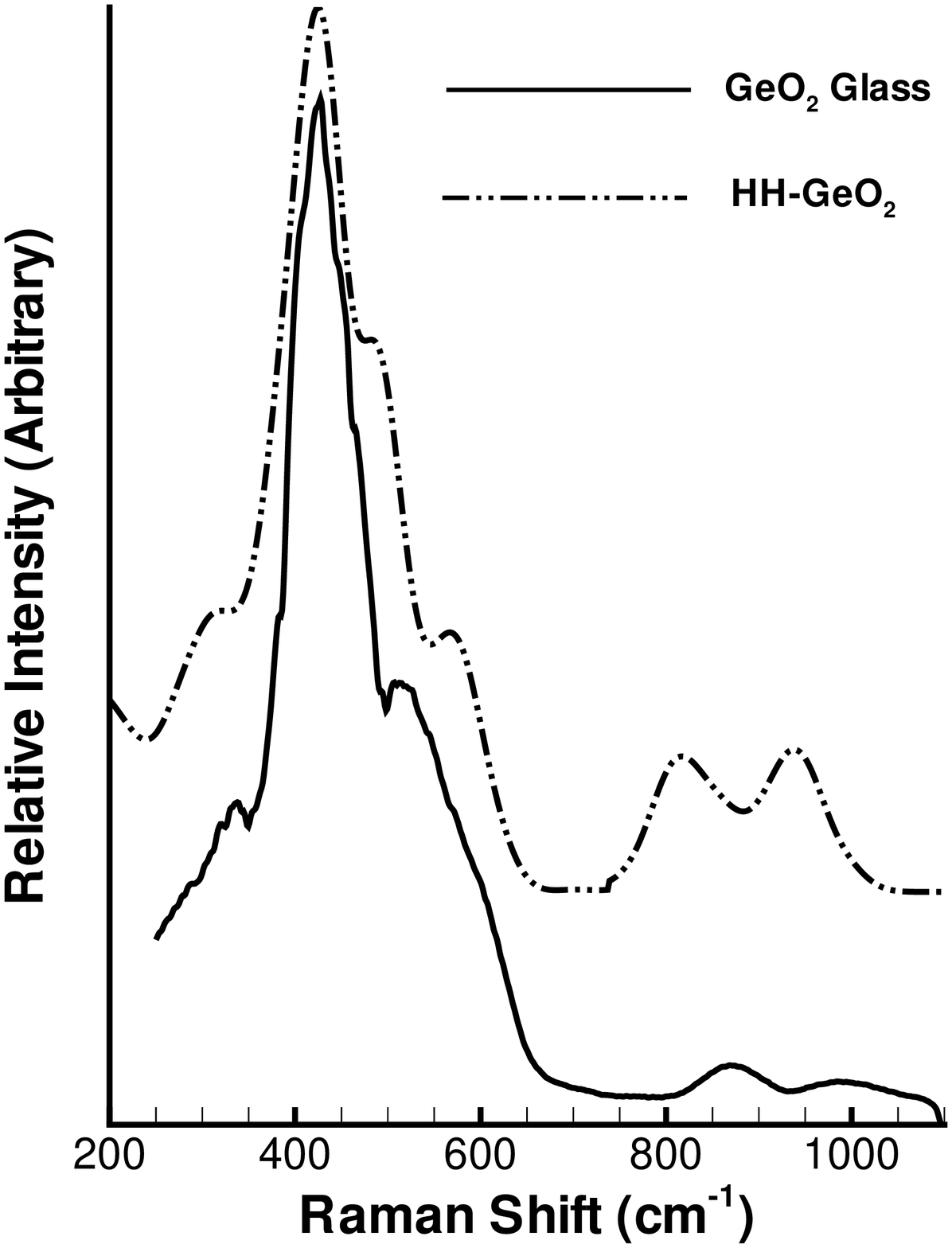}
\end{center}
\caption{a) Unpolarised Raman spectrum of $GeO_2$ glass showing the main vibrational bands, 
b) a Raman spectrum of a $Li_2O$-containing germanate glass showing the high frequency 
BO and NBO bands: The insert is a curve fit (deconvolution) of the high frequency envelope 
into its discrete vibrational bands (see table 2), c) Unpolarised Raman spectrum of 
$GeO_2$ glass (solid line) compared with calculated HH spectrum of $GeO_2$ glass 
(dashed-dot line) from Giacomazzi et al. \cite{Giacomazzi2005}. The calculated spectrum has 
been shifted so that the main vibrational band is coincident with the equivalent band of 
the experimental spectrum.}
\end{figure}

\par 

The  high frequency bands  observed  at $\simeq 860$ and $998~cm^{-1}$
are the TO  and LO split  asymmetric stretching bands of the  bridging
oxygens ($Ge-O-Ge$).  The $Ge-O-Ge$ bending  modes are observed in the
broad  region  between  $\simeq 500-620~cm^{-1}$   and have also  been
assigned  to TO ($\simeq  556~cm^{-1}$)  and LO ($\simeq  595~cm^{-1}$)
split   modes  associated   with significant   $Ge$  and   $O$  motion
\cite{Galeener1976}. A "{\em defect}"   band $D_2$  occurs  at $\simeq
520~cm^{-1}$.  This  defect mode  is the  equivalent  of the $D_2$ band
observed in $SiO_2$ glass  at $606~cm^{-1}$ and is assigned  similarly
to an   oxygen-breathing  mode associated   with  3-membered rings  of
$GeO_4$ tetrahedra and this assignment has  been recently supported by
the study of Giacomazzi et al. \cite{Giacomazzi2005} (Fig. 8c). 
The intensity of
this band  is much stronger relative  to the  main vibrational band at
$\simeq 420~cm^{-1}$ in comparison to the equivalent bands for $SiO_2$
glass.  This indicates that the  $GeO_2$ network, while being composed
predominantly of  6-membered rings of  $GeO_4$  tetrahedra (see above)
does have a larger proportion of 3-membered  rings relative to $SiO_2$
glass.  The  relatively  narrow band at   around  $420~cm^{-1}$ is the
symmetric stretching mode of the $Ge-O-Ge$ bridging oxygens. Its width
is much narrower than the equivalent band observed in $SiO_2$ glass at
$440~cm^{-1}$    and indicates  that  the   distribution  of $Ge-O-Ge$
intertetrahedral angles for $GeO_2$  glass  is narrower than that  for
$SiO_2$ glass consistent with the neutron and X-ray data above.

\par

\begin{table}
\begin{center}
\begin{tabular}{l|l}\hline
Frequency&Attribution \\ \hline
& \\
$\simeq60~cm^{-1}$&Boson peak; Acoustic mode? Related to glass fragility\\
& \\
$347~cm^{-1}$ (D1)&Ge "{\em deformation}" motion within the network\\
& \\
$420~cm^{-1}$&Symmetric stretching of bridging oxygens (BO) in 6-membered\\
 &$GeO_4$ rings: Ge-O-Ge\\
& \\
$520~cm^{-1}$ (D2)&"{\em Defect}" mode assigned to
breathing motion of bridging\\
&oxygens in 3-membered GeO4 rings\\
& \\
$500-620~cm^{-1}$&
Bending modes: Ge-O-Ge, TO ($556~cm^{-1}$) and LO split\\
&($595~cm^{-1}$)\\
& \\
$\simeq860~cm^{-1}$&TO asymmetric stretching of bridging
oxygens: Ge-O-Ge ($Q^4$)\\
& \\
$\simeq988~cm^{-1}$&LO asymmetric stretching of bridging
oxygens: Ge-O-Ge ($Q^4$)\\
& \\
Q species vibrations&NBO (non-bridging oxygen)
vibrations that occur upon\\
& addition of network modifiers such as alkalis\\
& \\
$\simeq 865~cm^{-1}$&$Q^3$ $GeO_4$ tetrahedra with 3 BO and 1 NBO \\
& \\
$780~cm^{-1}$&
$Q^2$ $GeO_4$ tetrahedra with 2 BO and 2 NBO\\
& \\
\hline
\end{tabular}
\end{center}
\caption{Raman band assignments for $GeO_2$ glass and for the different Q 
species observed upon addition of a network modifier such as an alkali or 
alkaline-earth cation.}
\end{table}

The origin of the  Boson peak (BP)   at $60~cm^{-1}$ (the peak  occurs
over a broad  range between $40-60~cm^{-1}$) remains controversial. It
has   been assigned   to   acoustic-like   harmonic modes,   localized
quasi-harmonic modes, and to the smallest  energy van Hove singularity
of    the crystal   (cf.   \cite{Pilla2003},  \cite{Hubbard2003}   and
references therein). Most recently there seems  to be a consensus that
the  origin  of the BP   is due to  optic-like  excitations related to
nearly rigid $SiO_4$  (or $GeO_4$) librations through hybridization of
the acoustic waves  \cite{Courtens2003}. It  exhibits a dependence  on
the fragility of the  glass (fragile glasses  have weak BP intensity),
as well as,  fictive temperature (for $SiO_2$  the BP shifts to higher
wavenumber with increasing fictive temperature). In addition, there is
a  monotonic frequency shift   in the BP  for  pure $SiO_2$,  to lower
wavenumbers with   the  addition of $GeO_2$   which  may indicate that
$GeO_2$ substitutes isomorphously into $SiO_2$ \cite{Hubbard2003}.

\par

The   Raman spectrum of    $GeO_2$  glass indicates   that the   T-O-T
intertetrahedral  angle and its  distribution are narrower for $GeO_2$
glass relative to $SiO_2$ glass, consistent with the X-ray and neutron
diffraction studies  (see above).  These latter  studies  also suggest
that    the medium-range structure   of   $GeO_2$  glass  consists  of
6-membered rings of GeO4 tetrahedra, similar to  those observed in the
$\alpha$-quartz polymorph of $GeO_2$, with a  high proportion of small
3-membered $GeO_4$ rings (relative  to $SiO_2$ glass).  An interesting
aspect  of the  medium-range  structure was   raised by  Henderson  et
al. \cite{Henderson1985}  and Henderson and Fleet \cite{Henderson1991}
using  Raman  spectroscopy.  They   suggested  that   the medium-range
structure of $GeO_2$ may actually consist of 4- rather than 6-membered
$GeO_4$ rings. This suggestion has not been explored further and there
have  not  been any X-ray   or neutron  scattering studies  that  have
compared $GeO_2$   glass  with   structures  containing  predominantly
4-membered     $GeO_4$   rings.       However,     Giacomazzi       et
al. \cite{Giacomazzi2005} recently used a model $GeO_2$ structure that
had exclusively 3 and 4-membered $GeO_4$ rings. Their model reproduced
the first sharp  diffraction   peak in  the neutron  static  structure
factor (indicative of medium  range  structure), and the  infrared and
Raman spectra  of $GeO_2$ glass   (Figure   8c) reasonably well.   The
question of whether or not the medium-range structure of $GeO_2$ glass
consists of 6-  or  4-membered rings  remains unanswered and  open for
further studies.

\subsection{Infra Red (IR) spectroscopy}

There  have been  relatively few  infra-red  studies of $GeO_2$  glass
primarily because the  IR spectra are  more difficult to interpret and
obtain than the  Raman spectra. One of the  earliest is that of Kaiser
et al. \cite{Kaiser1956} while more recent studies  have tended to use
IR in high-pressure studies \cite{Teredesai2005} for investigating the
onset of amorphous to amorphous phase transitions  (see below). The IR
spectrum of $GeO_2$ glass exhibits two peaks  at $560~cm^{-1}$ and one
at   $\simeq 870~cm^{-1}$ with a    shoulder at $\simeq  1000~cm^{-1}$
although the relative intensities for these two  bands are reversed in
the spectra of Galeener  et al. \cite{Galeener1983}. The low frequency
band  at $560~cm^{-1}$  is  the IR equivalent  of  the LO bending mode
observed in the Raman  spectrum at $\simeq595~cm^{-1}$ while the bands
at  $870~cm^{-1}$ and $\simeq 1000~cm^{-1}$ are   the IR equivalent TO
($870~cm^{-1}$) and  LO split  asymmetric stretching  of  the bridging
oxygens   \cite{Galeener1983}.       The    data   of     Galeener  et
al.   \cite{Galeener1983} also show  a   peak in their  IR reflectance
spectrum at $\simeq 340~cm^{-1}$   which   is the equivalent  of   the
$347~cm^{-1}$ Raman band. Galeener  et al. \cite{Galeener1983}  assign
this band however  to an LO   mode. In general the  LO  modes are more
intense in  the IR relative  to Raman spectra while  the  TO modes are
more intense in the Raman relative to IR spectra.

\par

With increasing  pressure,  the $560$ and  $870~cm^{-1}$ peaks broaden
and the region between   the bands ($\simeq 700~cm^{-1}$)  exhibits an
increase in  intensity  \cite{Teredesai2005}, although  part of   this
increase  is due  to  a  shift in   the $560~cm^{-1}$  band to  higher
wavenumbers with  increasing  pressure (up to $6~GPa$).   Teredesai et
al. \cite{Teredesai2005}   also  observe  with increasing   pressure a
decrease in wavenumber for both high  wavenumber bands. Above $6~GPa$,
all bands shift to higher wavenumbers coincident with the onset of the
pressure  induced   coordination  change of  $Ge$   noted   by Itié et
al. \cite{Itie1989}.   However glasses  decompressed  from   $9.5~GPa$
exhibit a $30~cm^{-1}$ red shift in the  position of the $870~cm^{-1}$
peak  with   no  shift   in    position of   the   $560~cm^{-1}$  peak
\cite{Teredesai2005}.

\subsection{Increasing Pressure and Temperature}

The effect  of pressure on  $GeO_2$  glass at ambient temperature  has
been investigated by Ishihara  et al. \cite{Ishihara1999} and  {\em in
situ}   by  Durben   and   Wolf   \cite{Durben1991}  and   Polsky   et
al.    \cite{Polsky1999}.    Up  to     $6~GPa$,    Durben and    Wolf
\cite{Durben1991} observe  a shift of the  main Raman band  at $\simeq
420~cm^{-1}$ to higher frequency  with concomitant broadening and loss
of  intensity. Between $6$ and  $13~GPa$ the main  Raman band broadens
and losses  intensity without a shift  in frequency. In  addition they
observe   the growth   of  a  broad   low  frequency  band  at $\simeq
240~cm^{-1}$  and  no  further  spectral changes are   observed beyond
$13~GPa$    up  to    $56~GPa$. However,   upon   decompression,   the
$520~cm^{-1}$    $D_2$ band characteristic   of   3-membered rings  is
enhanced   relative to uncompressed  $GeO_2$  glass and indicates that
3-membered   rings   are   formed  during   decompression   from  high
pressure.   Similar    results   were    obtained    by  Polsky     et
al.   \cite{Polsky1999} and    both    they  and  Durben    and   Wolf
\cite{Durben1991} observe subtle changes  in the Raman spectra between
$5$ and $10~GPa$ characteristic of the pressure induced change in $Ge$
coordination     observed by    {\em in-situ}    EXAFS   and  XANES  studies
\cite{Itie1989}.

\par

Up to 5  GPa, both  Durban and  Wolf  \cite{Durben1991} and Polsky  et
al. \cite{Polsky1999} suggest that  compression  of the $GeO_2$  glass
network is taken up be tetrahedral deformation with a smaller decrease
in the intertetrahedral angle. In  addition, they conclude that  there
is no increase in intensity of the $520 ~cm^{-1}$ $D_2$ band. However,
this   conclusion is questionable  given that  the  main Raman band at
$420~cm^{-1}$ appears to  move to  higher wavenumbers with  increasing
pressure  (above  $4~GPa$) and  as   Polsky  et al.  \cite{Polsky1999}
themselves  note, any   apparent   decrease in  the  intensity  of the
$520~cm^{-1}$ band   may simply be  a  consequence  of changes  in the
adjacent band at $420~cm^{-1}$. Examination of Figure  2 of Durben and
Wolf  \cite{Durben1991} shows that above $3.7~GPa$  the $420$ and $520
~cm^{-1}$  bands are merged  and  individual  bands are  unable to  be
discriminated. Below  $3.7~GPa$, the intensity  of the $D_2$ band also
cannot be determined without  some knowledge of  how the  spectra have
been normalized but a cursory examination appears to indicate that the
$D_2$ intensity  has increased  relative  to the  maximum in the  main
$420~cm^{-1}$ band.  Furthermore, Ishihara et al. \cite{Ishihara1999},
albeit using permanently densified $GeO_2$  glasses, note that  growth
of  the  $D_2$ band   correlates   with increasing  pressure;   higher
pressures  produce increased  $D_2$  intensity although  there  are no
permanent structural    changes for  glasses  decompressed  from below
$4~GPa$ \cite{Polsky1999}.

\par

High temperature  studies  have  been performed  by   Magruder III  et
al.  \cite{Magruder1987} and   Sharma et  al. \cite{Sharma1997}.  With
increasing  temperature Magruder III  et al. determined  that the high
frequency TO/LO split pair undergo a  2-fold loss of intensity between
$1723$  and  $2023~K$ and   that  the $D_2$  band  intensity   remains
constant. The   high frequency LO  band  at $988~cm^{-1}$  (figure 7a)
loses intensity     as the TO/LO  splitting   is  lost with increasing
temperature \cite{Sharma1997} but even in the melt phase two bands are
observed at  $\simeq 818$  and $940~cm^{-1}$,   respectively. However,
Sharma et  al. \cite{Sharma1997} observed an  increase in intensity of
the $D_2$ band and a shift of the main Raman  band at $420~cm^{-1}$ to
higher wavenumbers  combined with  a loss of  intensity while  the low
frequency   band at $347~cm^{-1}$ shifts  to  lower wavenumbers but is
observed up  to  $1623~K$. Both studies clearly  show  that  the Raman
bands observed  in $GeO_2$ glass remain even  in to the melt phase but
that there are subtle changes in intensities and band positions as the
glass is heated and eventually melts.
\par

\subsection{NMR spectroscopy}

The coordination environment of  $Ge$ in $GeO_2$ and alkali-containing
$GeO_2$ glasses  remains  an area of  intense  interest from  a  glass
perspective  because   of   the     unusual  physical properties    of
alkali-containing  germanate  glasses and the  possible   role of $Ge$
coordination     in        this      behaviour   \cite{Henderson1991},
\cite{Henderson2002}. $Ge$ NMR    would normally be  the technique  of
choice   to  investigate the    coordination  environment  of $Ge$  in
glasses. Germanium has  five naturally occurring isotopes  ($^{70}Ge$,
$^{72}Ge$, $^{73}Ge$, $^{74}Ge$ and $^{76}Ge$)  but only $^{73}Ge$  is
suitable   for NMR studies.  However,  while  $^{73}Ge$  NMR has  been
successfully   performed      on   solid    crystalline      compounds
\cite{Verkhovskii1999}, \cite{Verkhovskii2003}, \cite{Takeuchi2004} it
has  not  been   useful  for elucidating  the   structure of   glasses
\cite{Stebbins2002}, \cite{Du2006}.

\par
 
The  $^{17}O$   MAS    NMR spectra   of     $GeO_2$  glass, and    the
$\alpha$-quartz-like and rutile-like polymorphs of crystalline $GeO_2$
have been  obtained   by  Du and    Stebbins  \cite{Du2006}. The   two
crystalline polymorphs   and  $GeO_2$ glass   all   exhibit a   single
crystallographic  oxygen  site  similar to previous   data obtained at
lower    magnetic fields \cite{Hussin1998}.    The  oxygen site in the
$GeO_2$ glass is comparable to  that found in the $\alpha$-quartz-like
$GeO_2$ polymorph indicating  that the glass  consists of a network of
$GeO_4$  tetrahedra, consistent   with  X-ray and neutron   scattering
studies.

\section{Structure of densified liquid $GeO_2$}

Melting curves at  elevated   pressures were first reported  by
Jackson   \cite{Jackson1976}     in  the range     $1100-1700^oC$  and
$0.5-2~GPa$.  The high-temperature part of the  phase diagram was also
studied in
\cite{Brazhkin2003c} where an observed flatening of the melting curve at $P\simeq 2-4~GPa$ seems to be an indication of densification of the melt due to the
transformation of a quartz-like liquid into a rutile-like one.
\par Ordering of the melt structure in the same range of temperatures 
and pressures as
above was also reported \cite{Sharma1979} from  Raman scattering. 
Specifically, the
lowering of the  Raiyleigh line intensity from {\em in situ} high
pressure and temperature liquid Raman spectra was found to  be 
significantly lower than  for a glass quenched
at ambient pressure.  This suggest an  increased degree of short-range
order on compression   in the liquid and  a  more  ordered
network  structure.  However, it constitutes a major obstacle to studying
liquid $GeO_2$ at elevated pressures. Note that Moelcular Dynamics has 
not tried to simulate these experiments yet (see below).
 
\section{Structure of the binary $SiO_2$-$GeO_2$ glasses}

Germania and   silica  are prototype   glasses  for  continuous random
network   models,  based on  the corner  sharing   connection of their
$SiO_4$ and $GeO_4$ tetrahedra. The variations in the intertetrahedral
angles  and the presence   of  some structural defects (for   instance
dangling  bonds in $SiO_2$ glass)   allows  the formation of a   three
dimensional disordered network.

\par

Germanosilicate  glasses are widely   used as low-attenuation  optical
fibers,  yielding  numerous    studies on  their  physical   (optical)
properties  \cite{Duverger1998}.  Structural studies   are more scarce
despite   the need for  an understanding  of  the relationship between
glass properties and structure, particularly with respect to variation
of the local  site geometry, intertetrahedral angles,  ring statistics
and  their    relationship to   chemical   ordering, clustering and/or
substitution. A fundamental  question is to  determine  whether or not
germanosilicate glasses form a homogeneous network or if there is some
sort of clustering or phase separation.

\subsection{EXAFS and X-ray scattering}

An  early   $Ge$   K-edge  EXAFS investigation   \cite{Lapeyre1983} on
$12.5GeO_2-87.5SiO_2$  and $36.5GeO_2-63.5SiO_2$  glasses calculated a
$Ge-O$  distance of $1.73 ± 0.01~Å$  but  no second neighbors were
observed. A  more extensive study using a   combination of $Ge$ K-edge
X-ray absorption and wide  angle  X-ray scattering (WAXS)  experiments
were carried   out on  $GeO_2-SiO_2$  glasses containing $16-36~mol\%$
$GeO_2$  \cite{Greegor1987}. They  showed  that the  XANES spectra are
similar  with  increasing  $GeO_2$   content and  that   EXAFS-derived
distances are  $1.72 ± 0.02~Å$ for  $Ge-O$.  A $Ge$   coordination
number of $3.9 ± 0.2$, consistent with  $Ge$ in tetrahedral sites as
in  vitreous $GeO_2$,   and  a  mean   $Si-O$ distance of   $1.62~Å$
consistent  with   $Si$   remaining  tetrahedrally  coordinated,  were
obtained from their  WAXS data. These results  seem to be contradicted
by  a high  energy  X-ray  diffraction study  on  a  $29GeO_2-71SiO_2$
composition  glass \cite{Schlenz2003} that   found a mean coordination
number  for $Si$ and $Ge$  of $3.4 ± 0.05$. The authors explain this
low  coordination number  by proposing that  a  considerable number of
$Ge$ atoms are  connected with less  than four  oxygens or are  highly
distorted. Except for the latter study whose coordination number seems
questionable, all structural studies  are consistent with the presence
of $SiO_4$ and $GeO_4$ tetrahedra in binary $SiO_2-GeO_2$ glasses.

\par

The first  peak observed in the  X-ray radial distribution function is
at higher distance than would be expected assuming standard $Si-O$ and
$Ge-O$ distance  ($1.62$  and $1.72~Å$  respectively for tetrahedral
environment \cite{Greegor1987}). This suggests that the binary glasses
are not  a simple physical mixture of  $SiO_2$ and $GeO_2$  oxides. In
germanosilicate glasses, no   $GeO_2$ clusters are observed   and GeO4
tetrahedra are thus part of the  $SiO_2$ network. This is confirmed by
the second shell of  neighbors that has  been  observed in  EXAFS data
\cite{Greegor1987}. Indeed, this  peak corresponds to $Si$ and/or $Ge$
neighbors and both  its position and its  intensity  vary upon $Si/Ge$
substitution. $Ge$  atoms can thus  be accommodated within the $SiO_2$
network. Intertetrahedral angles were  calculated from EXAFS and  WAXS
data   and are between $139-149^o$,   which is  closer  to $144^o$ for
$SiO_2$ glass   (albeit  the   magnitude   of this     angle   remains
controversial, cf.   \cite{Henderson2005} than  $133^o$  for   $GeO_2$
glass. This suggests that at low $GeO_2$ content, the $Ge$ environment
is constrained by the silicate  network. These results are  consistent
with a  substitutional model in  which  $Ge$ substitutes randomly  for
$Si$   in the vitreous $SiO_2$   network with little  Ge clustering. A
random   substitution model is further   supported  by recent $^{17}O$
multiple quantum NMR  spectra  on $GeO_2-SiO_2$ binary  glasses  which
show  peaks  for  all   three types of  bridging  oxygens  ($Ge-O-Ge$,
$Ge-O-Si$, $Si-O-Si$), in proportions at least roughly consistent with
random mixing of the tetrahedral cations \cite{Du2006w}.

\par

The binary  $SiO_2-GeO_2$   glass  structure can  be  described  by  a
continuous random network of corner sharing $GeO_4$ and $SiO_4$ tetrahedra.

\subsection{Raman spectroscopy}

Information at the medium range  structure such as ring statistics and
the ordering of $Si$  and $Ge$ atoms, have  been primarily obtained by
Raman   spectroscopic      investigations         of   germanosilicate
glasses. Important modifications  appear between the Raman spectra  of
pure $GeO_2$  and $SiO_2$ and some  specific structures are present in
the spectra of the binary glasses.

\par
	
The band  at low  frequency shifts from  $437~cm^{-1}$  in $SiO_2$  to
$416~cm^{-1}$ in $GeO_2$ and  becomes sharper \cite{Sharma1984}.  This
band is attributed to the T-O-T (T=Si or Ge) symmetric stretching mode
and  is thus characteristic of the  distribution maximum  in the T-O-T
intertetrahedral  angles   \cite{Martinez2004}. Therefore,  it  can be
concluded that the fluctuation in the intertetrahedral angle decreases
as $GeO_2$ is introduced into the silica network.

\par

A complete Raman study from  pure $SiO_2$ to  pure $GeO_2$ was carried
out  by Sharma et al.  \cite{Sharma1984} in order  to characterize the
distribution of $SiO_4$ and $GeO_4$ tetrahedra. In the germanosilicate
glasses, a weak  band in the range  $970-1010~cm^{-1}$ appears that is
not present  in   pure  $SiO_2$  or $GeO_2$  glasses.   This  band  is
attributed  to the  antisymmetric  stretching motion  of  the bridging
oxygen of $Si-O-Ge$  linkages,  while the corresponding modes  for the
$Si-O-Si$ and $Ge-O-Ge$ linkages appear  at $\simeq 1110 ~cm^{-1}$ and
$\simeq  880~cm^{-1}$,  respectively. The position  of  the band is at
$\simeq 1000 ~cm^{-1}$ for  the $10GeO_2-90SiO_2$ glass  but decreases
to $\simeq 920 ~cm^{-1}$ for the  $90GeO_2-10SiO_2$ glass. This  shift
in position towards lower frequency is attributed to a decrease in the
$Si-O-Ge$ bond angle in the $GeO_2$-rich glasses \cite{Sharma1984}. In
the $50GeO_2-50SiO_2$ glass, the bands at $1100$ and $880~cm^{-1}$ are
stronger than the one  at $980 ~cm^{-1}$. This indicates the formation
of $Si-O-Ge$ bonds but  also the existence of  an important  number of
$Si-O-Si$ and  $Ge-O-Ge$  linkages.  According to these   authors, the
$Si/Ge$  ordering  is    likely non-ideal, which    supports  a random
distribution of $SiO_4$   and $GeO_4$  tetrahedra (see comment   above
regarding $^{17}O$  NMR).  In a  Molecular  Dynamics  simulation  of a
$50GeO_2-50SiO_2$   glass   \cite{Bernard2001},  a  large fraction  of
$Ge-O-Si$   bonds were  found,  as well   as, $Ge-O-Ge$ and  $Si-O-Si$
linkages. Based on the simulations,  Bernard et al. \cite{Bernard2001}
proposed that $Ge/Si$ ordering  occurred but  not  to the  extent that
phase  separation was  evident.   They also showed   that non-bridging
oxygens ($5\%$) were mainly localized in the $Ge$ environment.

\par

With a small addition of $GeO_2$, the $D_1$ and $D_2$ lines of $SiO_2$
glass  at $495$  and  $606~cm^{-1}$,  attributed to four-membered  and
three-membered rings of the $SiO_4$ tetrahedra in vitreous silica, are
still observed but the intensity  of the $D_1$ line decreases sharply,
while  that  of  the   $D_2$   lines decreases  slowly   and  broadens
\cite{Nian1989}.  Nian et  al.   \cite{Nian1989} suggested  that   the
substitution of $Ge$ for $Si$ in the vitreous $SiO_2$ network prevents
the  formation  of these  ring structures.  This was explained  by the
disruption of the fourfold and  threefold $SiO_4$ rings to accommodate
the    larger    $GeO_4$   tetrahedra  that   distort    the  silicate
network. Alternatively, the decrease  in intensity of $D_1$ and  $D_2$
lines could  also be due  to a change in  polarizability of the $Si-O$
bonds as  $Ge$ pulls electron density away  from $O$ attached to $Si$,
which  is an   explanation more  consistent with   the  preference for
3-membered  rings  in $GeO_2$   relative  to $SiO_2$  as indicated  by
diffraction  and  Raman   data (see  above).  With small   addition of
$GeO_2$, a new band appears at $710~cm^{-1}$ but its assignment is not
clear \cite{Nian1989}.   Above $15~mol\%$ $GeO_2$  content,  weak  and
broad shoulders are exhibited at $\simeq 568$ and $\simeq 670~cm^{-1}$
and    the band  at   $\simeq   800~cm^{-1}$  decreases  in  intensity
\cite{Sharma1984}.
	
\subsection{Evolution with pressure}

The  $Ge$    coordination   change  in   the   tetrahedral   framework
$SiO_2-GeO_2$ glasses is a reversible process that has to been studied
by  {\em  in situ}  high-pressure XAS measurements   at the  Ge K-edge
\cite{Majerus2004}. The pressure-composition  diagram in Fig. 9  shows
the  existence    of   three   regions  with   distinct    short-range
structures.  At low pressure,  the region corresponds to a tetrahedral
framework structure  (T domain), then,  an intermediate domain  with a
mixture of different sites, while, at higher pressure, the $Oc$ region
corresponds to   a  structure with  $^{[6]}Ge$. The  $^{[4]}Ge$   to $^{[6]}Ge$
transformation is a reversible with  an important hysteresis (a return
back  to the tetrahedral  site below $4~GPa$). The coordination change
is dependent on the  mean composition of  the glasses and extends over
higher-pressure   range  when  the  $SiO_2$    content  increases. The
remarkable dependence  of the $Ge$ coordination  change on the $SiO_2$
content  shows that the $Ge$  local structure is  strongly affected by
$Si$. The disruption of  the  $SiO_2$  tetrahedral network begins   at
$10~GPa$, as  evaluated by Raman spectroscopy \cite{Polsky1999}, which
is similar to the XAS data at  high $SiO_2$ content and indicates that
$Ge$   and   $Si$   convert    to   a  sixfold   coordination    state
simultaneously.   This  result suggests   that   the pressured-induced
transformations occur homogeneously  in the mixed  network  and may be
driven by the   oxygen atoms rather than by   the $Ge$  or  $Si$ atoms
\cite{Polsky1999}. Indeed, in such fully polymerized networks, oxygens
increase  their   coordination     from   two   to three     in    the
transformation. The transformation occurs  at higher pressure and over
a broader pressure range when the $SiO_2$ content increases. A careful
analysis of both XANES and EXAFS signals  supports a model of a mixing
of  $^{[4]}Ge$ and $^{[6]}Ge$ states in  the transition region, in agreement
with a kinetically hindered first-order process for the transformation
at room temperature.

\begin{figure}
\begin{center}
\includegraphics[width=0.65\linewidth]{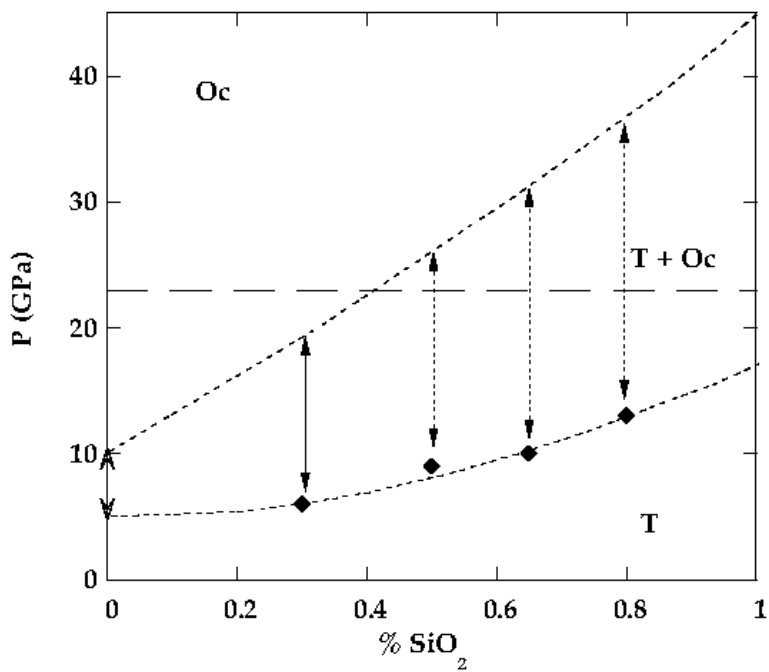}
\end{center}
\caption{Pressure-composition diagram of the $SiO_2-GeO_2$ glasses (after \cite{Majerus2004}) 
depicting the pressure-induced $Ge$ coordination change. The horizontal dashed line 
separates the pressure-range explored. Diamonds are the pressure onsets of the coordination 
change and the dashed curves delimit the intermediate domain separating the low-density 
form (T for tetrahedral) and high-density form ($Oc$ for octahedral).}
\end{figure}

\section{Molecular simulations and theoretical approaches}

\subsection{Force field parameters}

Several force field potentials  have been proposed to describe  either
the  crystalline phases of   $GeO_2$ or amorphous germania. All  these
potentials contain  a  long    range Coulombic  part, along    with  a
short-range repulsive term and an additional van der Waals-like term
\begin{eqnarray}
\label{V}
V_{ij}(r_{ij})&=&{\frac {Z_iZ_je^2}{r_{ij}}}+A_{ij}e^{-r_{ij}/ 
\rho_{ij}}-{\frac {C_{ij}}{r_{ij}^6}}
\end{eqnarray} 
where  $Z_i$ is  the charge on   ion $i$. The parameter
$\rho_{ij}$  serve  to determine the   steepness of   the short  range
repulsive  potential      and  is  known  as     the "{\em  softness}"
parameter. The parameters $A_{ij}$ and  $C_{ij}$  serve to adjust  the
positions  of  the first peak   in  each possible radial  distribution
function     to    experimental   findings.    Oeffner    and   Elliot
\cite{Oeffner1998}   have   fitted   equ. (\ref{V})   to   obtain cell
parameters,  density    and  elastic    constants  of   the   trigonal
$\alpha$-quartz    like    and    tetragonal  rutile-like   phases  of
$GeO_2$.  Bond angles  and  bond lengths  in   both the low   and high
pressure  phases are found  to  agree with  experimental findings. The
Raman and Infrared  vibrational spectra are  also simulated within the
harmonic  approximation  using the  bond-polarizability model  of Long
\cite{Long1953}.  Analysis from the   vibrational density of states of
$Ge-O-Ge$ motions   shows   that  for  $\alpha$-quartz-like   $GeO_2$,
symmetric and asymmetric bending motions are mostly confined to medium
and low frequency bands while symmetric stretching and bending motions
can be reasonably simulated at the anticipated frequencies.

\par

Matsui  and co-workers \cite{Tsuchiya2000} have used  the same kind of
approach,  i.e.  the fitting  of  equ. (\ref{V}), to  simulate another
structural phase transition,  namely the pressure induced  change from
$\alpha$-quartz-like $GeO_2$ to  rutile-like $GeO_2$  which happens at
$7.4~GPa$. The structure obtained at this pressure appears to be quite
similar to the    structure  calculated  for $SiO_2$  at    $21.5~GPa$
\cite{Bingelli1994}. Furthermore it is shown that $\alpha$-quartz-like
$GeO_2$ close to the transition  is  mechanically unstable as some  of
the  elastic moduli of  the lattice become negative. Specifically, the
decrease  of the  transverse elastic  constant   $C_{44}$ leads to  an
unstable shear that originates the  transformation to the  rutile-like
structure.  For increased pressures,  a post-rutile-like  structure is
found  \cite{Tsuchiya1998}  that has  a  $CaCl_2$-like structure which
consists of  tilted $GeO_6$  octahedra.  This  appears  to  be in
agreement  with   Brillouin  scattering results    of  $\alpha-GeO_2$ under
pressure
\cite{Grimsditch1998} which show that the shear constants are largely softened with respect to $SiO_2$ and can be related to shear instability.

\par

More recently, an alternative model has been proposed by Van Hoang
\cite{VanHoang2006a} for liquid  and amorphous germania  that is based
on a Morse-like  potential in a similar  manner to the potential given
by  Kim for $GeO_2$ \cite{Kim1996}.   We discuss below the  structural
predictions of the Van  Hoang potential. For completeness, we  mention
also the model potential proposed by Nanba \cite{Nanba1994} to account
for  $GeO_2-PbO-PbF_2$   glasses. However,  it    appears to show poor
agreement with the rutile-like properties of $GeO_2$.

\par

Topological  and   geometrical approaches    have also  been  proposed
\cite{Wefing1999}  in  order  to  generate continuous   random network
models of $GeO_2$ that reproduce  the experimental density, bond angle
distributions and neutron scattering  data \cite{Desa1988}. Araujo has
used  statistical mechanical techniques \cite{Araujo1996} to calculate
the   density of oxygen  vacancies   in  $GeO_2$  and  the  absorption
coefficient with respect to temperature.

\subsection{Simulation of liquid and amorphous germania}

\begin{figure}
\begin{center}
\includegraphics[width=0.6\linewidth]{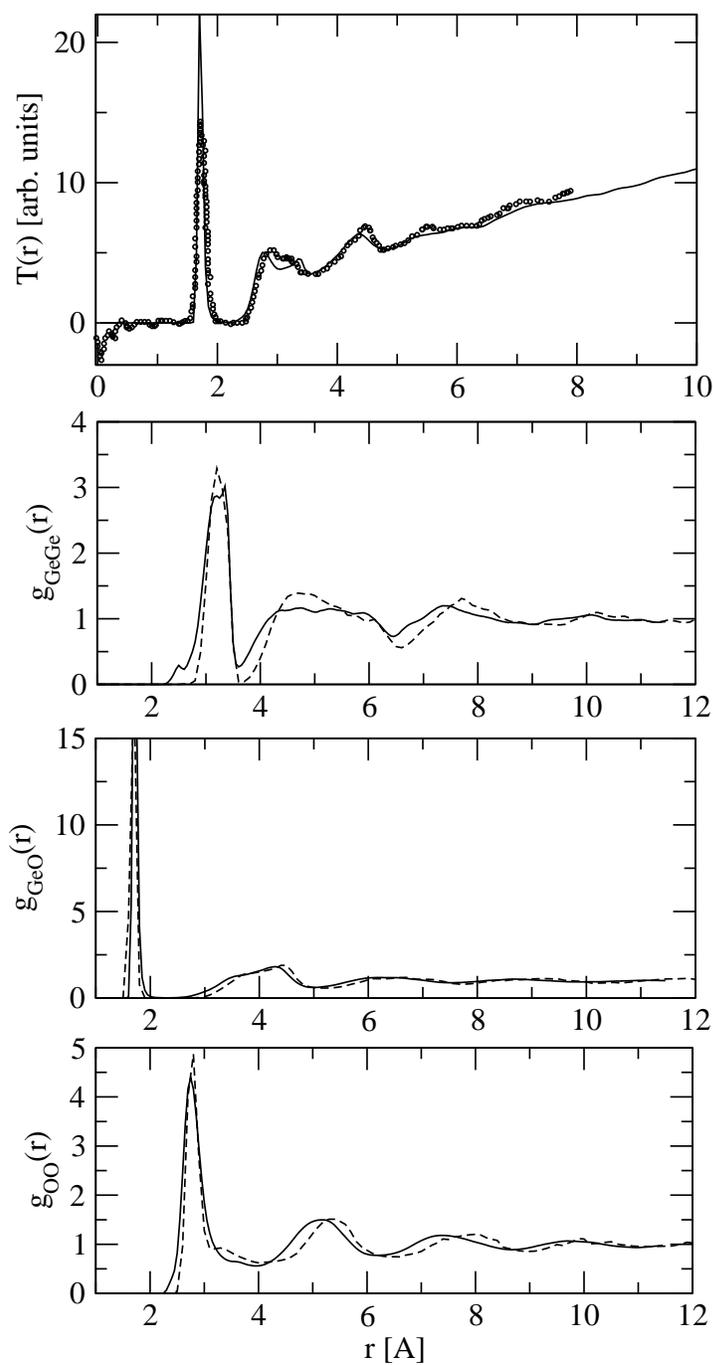}
\end{center}       
\caption{Upper panel: Simulated (dashed) Neutron structure \cite{Micoulaut2006} factor $T(r)$ 
compared to experimental findings \cite{Stone2001}. Lower panels: Partial structure factors 
of $300~K$ amorphous $GeO_2$ using the Oeffner-Elliot (solid line) \cite{Oeffner1998} and the 
Morse-like potential (broken line) \cite{VanHoang2006a}.}
\end{figure}

Most of the work  using the effective  potentials described above, has
been devoted to the  description of the  high temperature liquid where
experimental data is lacking. Gutierrez and Rogan \cite{Guttierez2004}
have  simulated  $GeO_2$  at   $1500~K$    and  $3000~K$.  At    these
temperatures, the   system  seems to be   made   of slightly distorted
$GeO_4$ tetrahedra  which are linked  by corners and have  a $Ge-O-Ge$
angle of  $130^o$, similar to  the experimental value in the amorphous
phase ($GeO_2$  glass). A volume  collapse, in the range $4-8~GPa$, is
seen  from the pressure-volume  curve  and may be   the signature of a
liquid-liquid   phase    transition,    in    analogy  with      water
\cite{Giovambattista2005}. Van Hoang has  carried out a similar  study
\cite{VanHoang2006b} under  the same kind  of conditions,  i.e. in the
high temperature  ($5000~K$)  liquid  with changing  density, and has
suggested that a diffusion maximum should be attained for a density of
about  $5~g/cm^3$. This anomaly appears  to be produced by competition
between the breakdown of  the tetrahedral network structure leading to
an increase in atomic  mobility, and the  packing effects arising from
densification that tend   to    reduce the mobility.    The  simulated
structure of  liquid $GeO_2$ and  $SiO_2$  appears to be  very similar
when  the partial atomic correlation   functions are properly rescaled
\cite{Micoulaut2004b}.

\par 
       
Micoulaut, Guissani  and Guillot  \cite{Micoulaut2006} have  used  the
Oeffner-Elliot  potential  to study the   glass and liquid  phases and
which allows comparison with experiments. In the glass, the structural
properties can   be   simulated relatively  well,  even   though  some
structural   limitations  of the   potential appear.   While the first
structural peak due to  $Ge-O$ interactions can  be modelled very well
at the expected distance  of $1.72~A$, as  can  the $O-O$ distance  at
$2.81~Å$,   the    $Ge-Ge$ correlations    appear   to  be  slightly
overestimated  ($3.32~Å$) with respect  to experimental values. This
overestimation  leads to    a  larger   calculated   value   for   the
intertetrahedral  angle than   that  obtained experimentally;  $159^o$
versus $130^o$,  respectively. It is  now well known that simple ionic
potentials such    as  the ones   reported   above \cite{Oeffner1998},
\cite{Tsuchiya1998} results in $Ge-O-Ge$  angles that are too wide,  a
situation that has been encountered and reported already for amorphous
silica  \cite{Rustad1991}.   However,   the  absence of    any $Ge-Ge$
interaction  in the effective Oeffner-Elliot  potential, except in the
Coulombic term,  may be  responsible  for the increased  distortion in
germania with respect to silica.  In spite of these deficiencies,  the
simulation correctly describes   the structure factor $S(Q)$  and  the
partial structure factors $S_{ij}(Q)$   (Figure 5) and allows one   to
infer the origin of the first sharp  diffraction peak (FSDP) as mostly
arising from $Ge-Ge$  correlations. However, overall  the potential is
found  to   reproduce  the features  of  neutron  scattering functions
(Figure  10)   reported   by   different   groups    \cite{Desa1988},
\cite{Stone2001}.

\par

Simulation using a  Morse-like potential \cite{VanHoang2006a} provides
a somewhat  better  agreement  with the   experimental  partial atomic
correlation functions (Figure 10) as the $Ge-Ge$  distance is found to
be    $3.21~Å$    at    $300~K$     with  correct     bond    angles
($\theta_{O-Ge-O}=108^o$  and $\theta_{Ge-O-Ge}=133^o$)  whereas  both
$Ge-O$ and $O-O$ distances are slightly underestimated ($1.69~Å$ and
$2.78~Å$ respectively) relative to experimentally derived values.

\subsection{Glass transition problem of strong glasses}

Enthalpy and glass transition temperature can be simulated rather well
\cite{Micoulaut2006} with   respect   to  calorimetric    measurements
\cite{Richet1990}.  With  the Oeffner-Elliot   potential, a  $T_g$  of
$900~K$,    is  found from  the  inflexion   point  of  the  potential
energy. This    value is close  to    the  experimental derived  $T_g$
($850~K$, \cite{Kiczenski2000}).  This appears to be rather unusual as
MD   simulations     on    similar    systems    \cite{Vashishta1989},
\cite{Vollmayr1996} predict much  higher glass transition temperatures
than the corresponding experimental ones. This is partially due to the
high quench rates applied. In the  present simulated systems, onset of
slow  dynamics at the nanosecond   scale occurs in  the  same range of
temperatures ($920~K$) which corroborate the calculated $T_g$ from the
inflexion point of the energy profile.

\par
    
When put  in contrast with  silica, a more  careful inspection  of the
self-diffusion coefficient $D$ with respect to the viscosity behaviour
\cite{Sipp2001} shows that   the agreement between  the simulated  and
experimentally measured $T_g$  reveals  an underlying failure  of  the
simulation technique.  The self-diffusion   coefficient D is  computed
from the mean-squared displacement of the  germanium and oxygen atoms,
and   shows   Arrhenius like  behaviour    $D=D_0\exp[E_i/T]$ at   low
temperatures,  whereas  at  higher    temperatures   ($T>1600~K$) some
curvature appears (Figure 11), similar to that found for molten silica
\cite{Horbach1999}. However,  the calculated oxygen diffusion constant
for $GeO_2$ at $1440~K$ is several orders of magnitude larger than the
reported data for  oxygen diffusion ($D_O=7× 10^{-14}~m2.s^{-1}$,
\cite{Tokuda1963}). A predicted   diffusion D constant from  viscosity
data $\eta$ using the  Eyring relation  $k_BT/ \eta D=\lambda$  (where
$\lambda$  is    a  hopping  length   of   about  several   Angstroms,
\cite{Glasstone1941}) shows that both silica and germania overestimate
the diffusion constants  with respect to  their simulated $T_g$'s thus
allowing  the  system to remain  in  a liquid-like  behaviour to lower
temperatures.  This underscores  both  the limitation of the  employed
potentials  and the size  of the  simulated   systems (actually  up to
several thousands atoms) to  accurately describe the  glass transition
of strong glass formers.

\begin{figure}
\begin{center}
\includegraphics[width=0.6\linewidth]{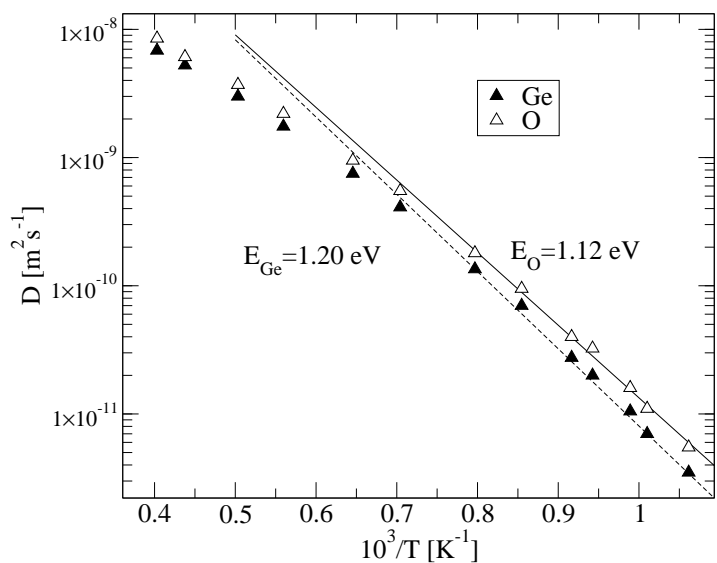}
\end{center}
\caption{Simulated diffusion for germanium and oxygen using the Oeffner-Elliot potential 
\cite{Micoulaut2006}.}
\end{figure}

\subsection{Equation of state}

The equation of state (EOS) of  $GeO_2$ has been reported by different
authors, either on  the basis  of simulations \cite{Guttierez2004}  or
from  empirical   models   based  on    simple   structural  arguments
\cite{Smith1995}. In the latter, Smith  and co-workers have shown that
a  two   state function,  taking   into  account  the   effect  of the
tetrahedral and   octahedral character at low  and  high pressures, is
able  to   describe the experimental  equation   of state  at $300~K$,
whereas molecular dynamics simulations  only succeed in simulating the
EOS in the low pressure range.

\par
 
At  higher  temperatures  and higher  densities,  Gutierrez  and Rogan
\cite{Guttierez2004} have shown  that  for  simulated $GeO_2$ in   the
$3.5-5.6~g/cm^3$   and  $T=1500-3000~K$  range,    pressure displays a
monotonic   decrease  with  molar  volume.   In   the same context,  a
Birch-Murnhagan   type   \cite{Birch1951} of     EOS  has   been  used
\cite{Micoulaut2006} to  fit a set  of $269$ simulated state points in
the  thermodynamic diagram. The   method allows the extraction  of the
isothermal compressibility $\kappa_T$ as a function of temperature and
density for density ranges lying between the ordinary glass density at
$300~K$        ($\rho=3.66~g/cm^3$,   \cite{Kamiya1986})   and    about
$2.5~g/cm^3$.  Progressive deviation  of the Birch-Murnhagan  EOS with
respect   to    the  simulated   thermodynamic  points     appear  for
$\rho<2.5~g/cm^3$ at high   temperatures.  At $2000~K$,  the  computed
compressibility    ($\kappa_T=9.13× 10^{-11}~Pa^{-1}$) is  rather
close to    the experimentally    measured    value of    Dingwell  et
al. ($\kappa_T=12.4× 10^{-11}~Pa^{-1}$, \cite{Dingwell1993}).

\par

Micoulaut  and    Guissani \cite{Micoulaut2006} have   used  a  Direct
Molecular Dynamics Method  \cite{Alejandre1995} to follow the equation
of  state   at zero pressure,  in  order  to predict  the liquid-vapor
coexistence  curve of germania on   the low (vapour)  and high density
(liquid) side in order to compare it  with experimental results in the
liquid up to  $1440~K$  \cite{Kamiya1986}.   Furthermore,  the  method
highlights  the quality  of  the effective potentials  employed at low
temperature. At  zero pressure and    low temperature ($300~K$),   the
density of a simulated Oeffner-Elliot $GeO_2$ glass \cite{Oeffner1998}
is  indeed  $3.70~g/cm^3$ whereas the density   of a simulated $GeO_2$
glass using an alternative potential \cite{Tsuchiya1998} substantially
disagrees with the experimental low temperature  density of the liquid
($\rho=4.25~g/cm^3$     as   compared    to         the   experimental
$\rho=3.66~g/cm^3$). Note however that this potential was used to study 
pressure induced rigidity in $GeO_2$ (see below, \cite{Trachenko2004})
and the density at zero pressure was found \cite{Trachenko2006}
to be $3.9~g/cm^3$, i.e. much closer to the experimental value. The thermal
history of the simulation appears therefore to be crucial in this case.
\par 
Using a Wegner type expansion \cite{Wegner1972},  a critical point for
germania   is   predicted   and     is   located    at   $T_c=3658~K$,
$\rho_c=0.59~g/cm^3$ and  $P_c=40~MPa$ \cite{Micoulautnp}. For  the
Tsuchiya potential  \cite{Tsuchiya1998}, the location  of the critical
point seems to be much higher in temperature \cite{Guissani2006}. This
shift may  arise   from the increased  charges  used  in the effective
potential.
\subsection{Pressurized germania}

The application of pressure to amorphous  germania seems to affect the
structure   stepwise. Experimentally  a  jump  in  bond distance  from
$1.72~Å$ to $1.86~Å$ is observed at around $9~GPa$, signalling the
conversion of  tetrahedral  to octahedral  local structure as  already
described.  However, numerical simulations show \cite{Micoulaut2004a},
at least in the  low pressure range,  that this conversion is somewhat
more subtle. For pressures up  to $2~GPa$, long-range correlations are
reduced, as seen from the shift to  higher wave vector of the position
of the FSDP; similar to experimental observations \cite{Sugai1996b}. In
addition, a  reduction is observed  in the intertetrahedral bond angle
($Ge-O-Ge$)  and then for $P=3~GPa$ a  sharp distortion of the $GeO_4$
tetrahedron   occurs (Figure 12).  These results  are accompanied by a
global increase in the number of oxygen neighbours  in the vicinity of
a     germanium atom  that     parallel   the   increase  in   density
\cite{Micoulaut2006}.
\begin{figure}
\begin{center}
\includegraphics[width=0.8\linewidth]{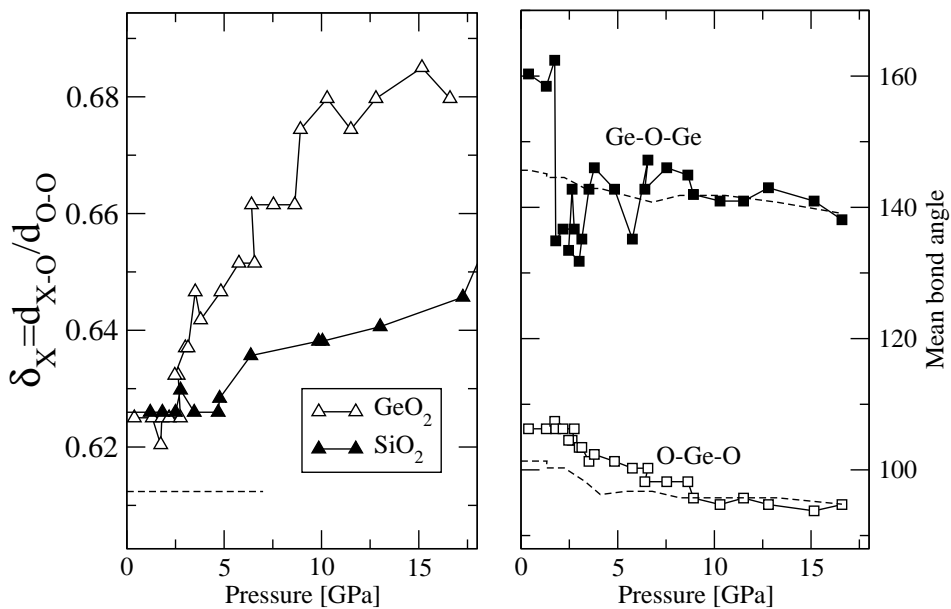}
\end{center}
\caption{Local structure of germania with applied pressure \cite{Micoulaut2004a}. 
Right panel: Distortion parameter $\delta_X$ of regular $GeO_4$ and $SiO_4$ tetrahedra, 
as a function of applied pressure ($X=Ge$, open triangles). For comparison, the same 
parameter for $SiO_2$ ($X=Si$, filled triangles) is shown. The broken horizontal line 
represents the value of the perfect tetrahedron $\delta=\sqrt{3/8}$. Right panel: 
Mean bond angles $Ge-O-Ge$ and $O-Ge-O$ with respect to compression (open and filled symbols) 
and decompression (broken curves).}
\end{figure}

\par

Sharma and  co-workers  \cite{Shanavas2006}  have   studied both   the
pressure induced   structural  changes  of  the   $\alpha$-quartz-like
$GeO_2$  polymorph  and  amorphous  $GeO_2$ using  the Oeffner-Elliott
potential  in the  (N,P,T) ensemble.  The results show   that both the
average  bond  distance ($Ge-O$) and  the  average  Ge coordination in
$\alpha$-quartz-like $GeO_2$ undergo  a sharp change at around $8~GPa$
under  compression, similar to the   experimental findings of Itié  et
al. \cite{Itie1989}.   On decompression, the  denser phase  transforms
back to a  lower-density phase at  $\simeq 2~GPa$. The details of  the
number of oxygen neighbours around a $Ge$ atom shows however, that the
high density phase is not fully six-fold  coordinated, as about $15\%$
five-fold    and  $20\%$ four-fold   germanium can    be found (Figure
13). Less abrupt changes are expected for vitreous $GeO_2$ (Figure 13)
where   a  majority of  six-fold   germanium only occurs for pressures
larger than $20~GPa$.

\par

The structural    changes with pressure are  more   dramatic in liquid
($1650~K$) $GeO_2$  as a sudden loss  of five-fold germanium atoms and
an almost  six-fold  coordinated structure is obtained   for pressures
larger than $12~GPa$.

\begin{figure}
\begin{center}   
\includegraphics[width=0.7\linewidth]{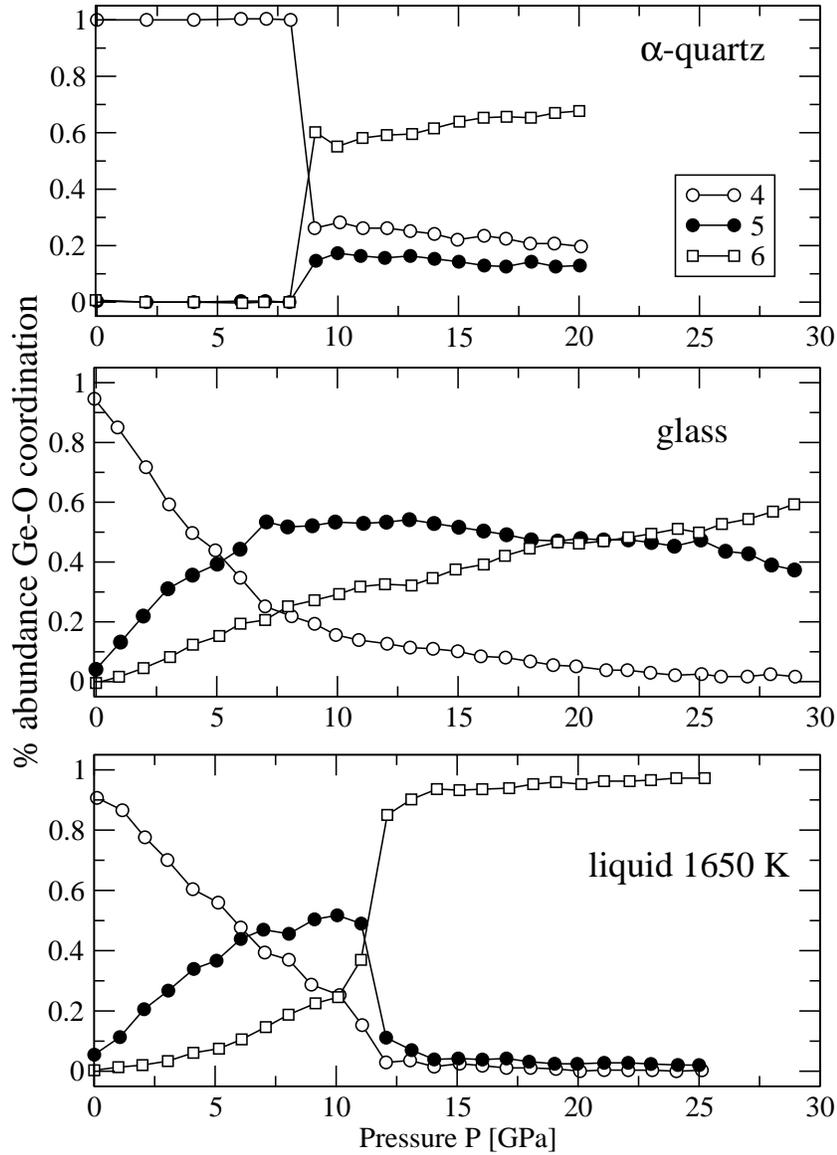}    
\end{center}
\caption{Calculated variation with pressure \cite{Shanavas2006} of the 
fractional 
abundance of $Ge$ coordination in the $\alpha$-quartz-like $GeO_2$ 
polymorph, glassy 
($300~K$) and liquid $GeO_2$ ($1650~K$).}
\end{figure} 
Finally, it appears that  the evolution of the intermediate range
order   with pressure or  density is  selective   as MD simulated ring
statistics
\cite{Micoulaut2007} show that rings with more than six germania tetrahdera tend 
to  disappear for densities larger than  $5~g/cm^3$ whereas the growth
of edge-sharing $GeO_6$  octahedra signals a behaviour similar
to $TiO_2$.
\subsection{Pressure induced rigidity and intermediate phases}

Trachenko et al.  \cite{Trachenko1998},  \cite{Trachenko2000} have
been  investigating the network rigidity  of $GeO_2$ and $SiO_2$ under
pressure.  Rigidity  usually appears  when  the  number  of mechanical
constraints per  atom,  arising  from  interatomic interaction (mostly
bond  stretching and bond bending)  becomes greater than the number of
degrees of freedom \cite{Rigidity1999}.   In network glasses,  this is
generally  achieved by the addition of  cross-linking elements such as
germanium into a   basic flexible structure  containing e.g.  selenium
chains.  This  leads to an  increase  of the network mean coordination
number $\bar r$  (and to the  increase of constraints) and produces  a
stiffening  of   the  structure  and  ultimately  a   floppy  to rigid
transition. The onset of rigidity  and the way  it percolates has been
documented for various glass-forming systems. In recent years however,
a reversibility window   \cite{Selvenathan2000},  \cite{Boolchand2000}
has been discovered  located between the floppy  and rigid phases, and
which manifests itself by the loss of irreversibility (and hysteresis)
of the heat flow when cycling through the glass transition temperature
region. A similar state  can  be found  in glassy $GeO_2$  and $SiO_2$
under pressure.
       
\begin{figure}
\begin{center}
\includegraphics[width=0.6\linewidth]{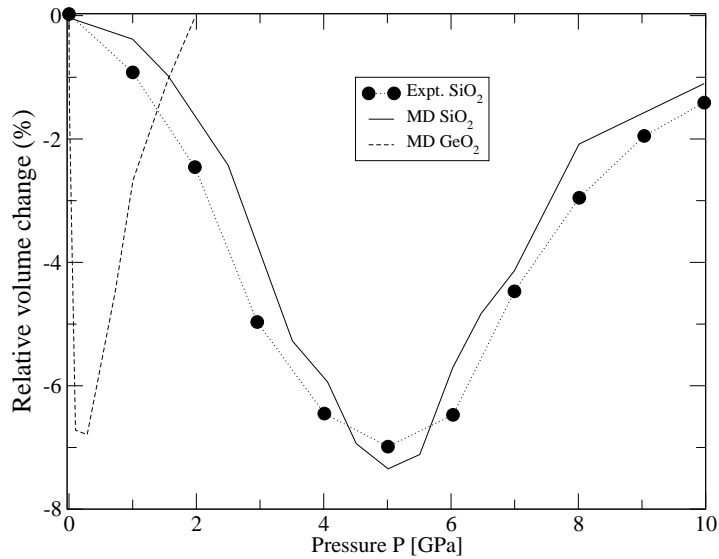}
\end{center}
\caption{Relative variation in volume in $GeO_2$ and $SiO_2$ glasses under pressure compared 
with experiments on $SiO_2$ \cite{Trachenko2004}.}
\end{figure}
      
\par
 
Pressure induced rigidity  in $GeO_2$ glass  using MD  simulations has
been  addressed recently  \cite{Trachenko2004}.  Here the increase  in
connectivity (or mean coordination number  $\bar r$) is achieved  with
the increase of the glass density or  the application of pressure that
produces a  tetrahedral    to  octahedral conversion.  Thus   pressure
introduces locally rigid   higher-coordinated  units in an   otherwise
flexible tetrahedral network of $GeO_4$ tetrahedra. Densification with
temperature  under  pressure  can  take  place  in a  pressure window,
centred around the rigidity percolation transition. The density change
is  about $7\%$. This new effect  has been rather  well documented for
silica  \cite{Trachenko1998},    \cite{Trachenko2000}   and   compared
successfully    with     experimental    results,
\cite{Tsiok1998} and  additional  simulations showing the  loss of low
frequency modes  in the effective  vibrational densities of  states at
the same  pressure  where densification  occurs.  However, it has only
been  shown that  the pressure  window in germania  is centred  around
$0.5~GPa$, i.e. considerably lower  than for silica ($5~GPa$)  (Figure
14). However, it is another signature of  the increased sensitivity to
pressure change of  $GeO_2$  with   respect to $SiO_2$.  Indeed,   the
tetrahedral to octahedral conversion  of amorphous  $SiO_2$ manifested
by the jump in $Si-O$  bond distances is  found to be around  $13~GPa$
\cite{Polian1990} whereas the same jump is  found to be at $8~GPa$ for
$GeO_2$ glass \cite{Itie1989}.  It  is therefore not surprising at all
that onset of rigidity manifests at lower pressures in $GeO_2$.

\subsection{{\em Ab initio} studies of c-$GeO_2$ and germania}

One  way to  circumvent the possible  failures of  the above mentioned
semi-empirical potentials, is the use of {\em ab initio} methods, especially
under  extreme conditions where   the  potentials are  not necessarily
reliable.  Hafner and co-workers  \cite{Lodziana2001} have studied the
high-pressure transformations up to  $70~GPa$ of  crystalline $GeO_2$,
using density functional theory  with a pseudopotential method,  and a
local density  approximation. It  appears  from this  computation that
several    high-pressure  phases can  exist in   $GeO_2$   which are a
tetragonal    $CaCl_2$ type at    $40~GPa$,  an  $alpha-PbO_2$-type at
$40~GPa$ and finally  a pyrite-type  crystal  at $70~GPa$, similar  to
those  observed    experimentally  (see   above)   (Figure 3).   These
transformations highlight the analogy of the phase transition sequence
between $SiO_2$ and  $GeO_2$ polymorphs  at high pressure.  Additional
studies concerning the  electronic properties of these polymorphs have
been reported by Christie et  al. \cite{Christie2000}, using the  same
tools.  This allows determination of  the lattice parameters, cohesive
energy  and bulk   modulus  by minimizing   the  total  energy  of the
solid.  In addition, an equation  of  state for the  polymorphs can be
fitted     with     a     Birch-Murnhagan    EOS     \cite{Birch1951},
\cite{Murnhagan1944} or the density of states.

\par
 
{\em Ab initio} studies of  amorphous germania have been only reported
recently  \cite{Giacomazzi2005}  using the  same  numerical scheme but
with an improved  density  approximation (generalized gradient).  This
enables determination of the neutron structure  factor (Figure 4), the
infrared and the Raman spectra, all  of which show good agreement with
experimentally derived data (cf.,  Figure 8c). It furthermore provides
insight into the 3-membered ring distribution  and the so-called $D_2$
line first described by Galeener  and workers \cite{Galeener1983}. The
projection   of   vibrational eigenmodes  onto   natural  or  isotopic
substituted oxygen breathing motions in these rings shows that a broad
peak centered at $520~cm^{-1}$ and corresponding to the experiment, is
blueshifted  with $^{18}O$ by  $26~cm^{-1}$. The number of these rings
is found  to be about $20\%$  of the oxygen  atoms. On the other hand,
similar calculations do   not   seem to  support   the  assignment  of
four-membered rings to the $D_1$ line found at $347~cm^{-1}$. Instead,
this band seems to arise from coupled motions of $Ge$ and $O$ atoms.

\section{Summary and Conclusions}
 Studies  on the structure of crystalline,  liquid  and glassy $GeO_2$
 continue to be of interest to a number of  researchers in physics and
 glass,  materials and geological  sciences.  This breadth of interest
 stems  from the fact that while  there are close similarities between
 $GeO_2$  and $SiO_2$, there are also  distinct differences which make
 $GeO_2$ useful  as  an   analogue for   studying the  high   pressure
 behaviour of oxide glasses.   Crystalline $GeO_2$ polymorphs  behave,
 with increasing  temperature  and pressure, in  a manner   similar to
 crystalline  $SiO_2$   polymorphs. However,   pressure induced  phase
 transformations   generally occur  at    much lower  pressures   than
 equivalent $SiO_2$ phases.  This  is   because the   larger   $GeO_4$
 tetrahedron (relative to $SiO_4$ tetrahedron)  is more distorted  due
 to greater variability in the $O-Ge-O$ angles. This  makes the use of
 $GeO_2$ polymorphs attractive as  $SiO_2$ analogues in  high pressure
 studies  for studying  possible pressure induced  structural changes,
 since the pressure ranges  required are much more accessible. 
\par
$GeO_2$ glass has also been   considered as being somewhat similar  to
$SiO_2$ glass.  The first three interatomic distances in the glass are
reasonably  well resolved and  indicate that, like  $SiO_2$ glass, the
network is composed of tetrahedra linked together through their corner
bridging oxygens.  However,  there are significant differences between
the   two  glass networks.  $GeO_2$   glass  has a  much  smaller mean
$Ge-O-Ge$  angle  and a much  higher  proportion  of 3-membered rings,
relative  to $SiO_2$ glass.  Furthermore,  there may be differences in
the   intermediate-range structure with   $GeO_2$ glass possibly being
composed  of  4- membered rings,   rather than the  currently accepted
6-membered rings similar  to $SiO_2$ glass.  In  addition, application
of high pressure   readily converts  4-fold Ge  to  6-fold Ge,  via  a
transitionary 5-fold coordination, at much  lower pressures than found
for $SiO_2$  glass.    In  the liquid  state  $GeO_2$   retains 4-fold
geometry to high  temperature   but with broadened  $Ge-O-Ge$   angles
although   numerical studies of  liquid   GeO2 indicate that while the
$GeO_4$ tetrahedra are  distorted, the $Ge-O-Ge$ angle remains similar
to that found in  the $\alpha$-quartz-like $GeO_2$ polymorph.  Furthermore,
simulation  of  pressure effects   indicate that  the pressure-induced
transformation  from 4- to   6-fold Ge observed  experimentally at  $\simeq
9~GPa$, may be quite subtle. Onset  of the simulated transition occurs
at $2~GPa$ with  loss of long-range  correlations, a reduction  in the
$Ge-O-Ge$  angle,  followed  by  a sharp  distortion   of the  $GeO_4$
tetrahedra at $3~GPa$. This is accompanied by the onset of rigidity at
much  lower pressure   than   observed for   $SiO_2$  glass.  When  Ge
substitutes for  $Si$  along the  $GeO_2-SiO_2$   binary, there  is no
evidence for  clustering or phase  separation of the glass network and
it is composed of $SiO_4$ and $GeO_4$  tetrahedra. The substitution is
random  with no  heterogeneity    induced  in the  combined   network.
However, with increasing  pressure Ge undergoes  a coordination change
from 4-  to 6-fold coordination. The pressure  at which this occurs is
dependent  upon the $SiO_2$    composition indicating that  Si  has an
influence on the local structure of Ge.  In addition, there is a broad
pressure-composition range  over  which Ge is in   both 4-  and 6-fold
coordination.

\section*{Acknowledgements}
GSH acknowledges  funding  from NSERC  via a Discovery  grant. He also
thanks his  co-authors, the  Institut  de Minéralogie et  Physique des
Milieux Condensés (IMPMC), and  the CNRS for a  very enjoyable stay at
IMPMC. The   authors acknowledge discussions  and correspondance  with
P.  Boolchand,    G. Calas,  M.T.  Dove,   S.R.  Elliott,  G.  Ferlat,
L.  Giacomazzi, B. Guillot,   Y.  Guissani, G. Hovis,  A. Pasquarello,
J.C. Phillips, D.L. Price, P. Richet, J.F. Stebbins and K. Trachenko.
$$ $$

\end{document}